\documentclass[a4,12pt,eclepsf]{article}
\usepackage[dvips]{graphicx}
\usepackage{amsmath, amssymb, comment, bm, fancyhdr}

\newcommand{\BM}{\begin{minipage}}
\newcommand{\EM}{\end{minipage}}

\newcommand{\wt}[1]{\widetilde{#1}}

\begin{document}
\renewcommand{\theequation}{\thesection.\arabic{equation}}
\thispagestyle{empty}
\vspace*{20mm} 
\begin{center}
{\LARGE {\bf Chiral Symmetry Restoration in}} \\[5mm]
{\LARGE {\bf Holographic Noncommutative QCD}} 
\vspace*{30mm}

\renewcommand{\thefootnote}{\fnsymbol{footnote}}

{\Large Tadahito NAKAJIMA, \footnotemark[1]\,
Yukiko OHTAKE \footnotemark[2]
and \,Kenji SUZUKI \footnotemark[3]} \\
\vspace*{20mm}

$\footnotemark[1]$ {\it College of Engineering, Nihon University, 
Fukushima 963-8642, Japan} \\[2mm]
$\footnotemark[2]$
{\it Toyama National College of Technology, Toyama 939-8630, Japan} \\[2mm]
$\footnotemark[3]$
{\it Department of Physics, Ochanomizu University, 
Tokyo 112-8610, Japan} \\[30mm]

{\bf Abstract} \\

\end{center}

We consider the noncommutative deformation of the Sakai--Sugimoto model at 
finite temperature and finite baryon chemical potential. The space 
noncommutativity is possible to have an influence on the flavor dynamics of 
the QCD. The critical temperature and critical value of the chemical 
potential are modified by the space noncommutativity. The influence of the 
space noncommutativity on the flavor dynamics of the QCD is caused by the 
Wess--Zumino term in the effective action of the D8-branes. The intermediate 
temperature phase, in which the gluons deconfine but the chiral symmetry 
remains broken, is easy to be realized in some region 
of the noncommutativity parameter. 


\clearpage

%

\section{Introduction}
\setcounter{page}{1}

Noncommutative gauge theories (gauge theories on noncommutative space) 
naturally arise as low energy theories of D-branes in 
Neveu-Schwarz-Neveu-Schwarz (NS-NS) $B$-field background 
\cite{CDS, DH, AASJ, SJ, SW}. The space noncommutativity brings nontrivial 
properties on the gauge field theory at the quantum level. A remarkable 
example is the mixing between the infrared and the ultraviolet degrees of 
freedom. Although the product of the momentum and 
the noncommutativity parameter plays the role of ultraviolet cut-off, the 
result is singular when the momentum or the noncommutativity parameter is taken to zero. This phenomenon is referred to as the UV/IR mixing \cite{MRS}. Although the noncommutative gauge theories have been studied extensively, it is hard to investigate them in the perturbative approach. Little is known of the 
non-parturbative information of noncommutative gauge theories.

The noncommutative gauge theories have gravity duals whose near horizon region 
describes the large $N$ limit of the noncommutative gauge theories. This is 
first constructed by Hashimoto and Itzhaki \cite{HI}, and Maldacena and Russo 
\cite{MR}. Using the gravity dual description, we can investigate the 
non-perturbative aspects of the large $N$ limit of the noncommutative gauge 
theories. For instance, the space noncommutativity modifies the Wilson loop behavior \cite{dhar_kita, lee_sin, taka_naka-suzu} and glueball mass spectra 
\cite{NST}. The gravity duals of noncommutative gauge theories with matter in 
the fundamental representation have also been constructed by adding probe 
flavor branes \cite{APR}. Throughout the construction of gravity duals of 
noncommutative gauge theories with matter degrees of freedom it has been found 
the space noncommutativity is also reflected in the flavor dynamics. For 
instance, the mass spectrum of mesons can be modified by the space 
noncommutativity \cite{APR}.

The gauge/gravity correspondence can be developed to study the physics of chiral symmetry breaking, besides confinement in low energy behaviors of quantum 
chromodynamics (QCD) \cite{EEKT, M, PZ}. An attempt to describe the low energy 
hadron physics in terms of the gauge/gravity correspondence is called the 
holographic QCD model. The gravity duals of theories with non-Abelian chiral 
symmetry have been constructed by Sakai and Sugimoto \cite{SS1, SS2}. This 
model, so called Sakai--Sugimoto model, is known as one of the most accurate 
holographic realizations of the real QCD. The holographic QCD models can be 
modified to introduce a finite temperature and a finite baryon chemical 
potential. In the Sakai--Sugimoto model, phase transition of chiral symmetry 
restoration has been investigated at finite temperature in 
\cite{ASY, PS, PSZ, KSZ} and finite baryon chemical potential in 
\cite{HT, PS2, Y}. The phase diagrams of holographic QCD have been shown to 
contain a line of first order phase transitions.

The presence of magnetic field promotes the spontaneous chiral symmetry 
breaking in QCD. The effect of an external constant $B$-field in the 
Sakai--Sugimoto model on the phase structure of the QCD has been 
investigated and it has been shown that the external constant $B$-field 
promotes chiral symmetry breaking in the QCD \cite{BLL, JK, H}. This fact 
seems to suggest that NS-NS $B$-field background, that causes the space 
noncommutativity in the QCD, has a certain effect on the chiral phase 
transition of the QCD.

There have been some papers on how to influence the NS-NS $B$-field background 
on the strong coupled dynamics of the QCD in the holographic approach. 
The influence of the NS-NS $B$-field background on the drag force in thermal 
plasma of Yang-Mills theories has been investigated in \cite{MTW, PR, SP}. 
The influence of the NS-NS $B$-field background on the chiral symmetry 
restoration in the QCD at finite temperature has also been investigated 
in the manner of adding a constant background magnetic field besides the NS-NS 
$B$-field in Sakai--Sugimoto model \cite{SSX}.

In this paper, we construct the noncommutative deformation of the 
Sakai--Sugimoto model in imitation of the method by Arean et al. \cite{APR} 
and investigate the influence of the NS-NS $B$-field background on the chiral 
symmetry restoration in the QCD at finite temperature and finite baryon 
chemical potential. As will be seen later, the introduction of the 
Wess--Zumino (WZ) term produces the dependence of the effective action of the 
probe D8-brane on the NS-NS $B$-field background. The existence of the 
gauge fields in the total effective action of the probe D8-brane carries the 
dependence of the baryon chemical potential on the QCD.

This paper is organized as follows. In section 2, we construct the 
noncommutative deformation of the holographic QCD model at finite temperature 
and finite baryon chemical potential. In section 3, we investigate the chiral 
phase transition in the noncommutative QCD within the framework of the noncommutative deformation of the holographic QCD model. Section 4 is devoted to 
conclusions and discussions.

%
%

\section{Noncommutative deformation of the holographic QCD model at finite 
temperature and finite baryon chemical potential} 

\setcounter{equation}{0}
\addtocounter{enumi}{1}

In this section, we consider a noncommutative deformation of the holographic 
QCD (Sakai--Sugimoto) model at finite temperature based on the prescription 
of Arean--Paredes--Ramallo \cite{APR}. 
The holographic QCD model is a gravity dual for a $4+1$ dimensional QCD with 
${\rm U}({\rm N}_{f})_{L} \times {\rm U}({\rm N}_{f})_{R}$ global chiral 
symmetry whose symmetry is spontaneously broken \cite{SS1, SS2}. This model is 
a $\text{D4-D8-}\overline{\text{D8}}\text{-}$brane system consisting $S^{1}$ 
compactified $N_{c}$ D4-branes and $N_{f}$ $\text{D8-}\overline{\text{D8}}$ 
pairs transverse to the $S^{1}$. The near-horizon limit of the set of $N_{c}$ 
D4-branes solution compactified on $S^{1}$ takes the form
\begin{align}
ds^{2} &=\left(\frac{U}{R_{\rm D4}}\right)^{3/2}\Bigl(-(dt)^{2}+(dx^{1})^{2}
+ (dx^{2})^{2}+(dx^{3})^{2} + f(U)\,d\tau^{2} \Bigr) \nonumber \\
& + \left(\frac{R_{\rm D4}}{U}\right)^{3/2} 
\left(\frac{dU^{2}}{f(U)}+U^{2}d\Omega_{4}^{2} \right)\,,\nonumber \\
& R_{\rm D4}^{3}=\pi g_{s}N_{c}l^{3}_{s}\,, \qquad 
f(U)=1-\frac{U_{\rm KK}{}^{3}}{U^{3}}\,,
\label{201}
\end{align}
where $U_{\rm KK}$ is a parameter, $U$ is the radial direction bounded from below by $U \geq U_{\rm KK}$, $\tau$ is compactified direction of the D4-brane world volume which is transverse to the $\text{D8-}\overline{\text{D8}}$-branes, $g_{s}$ and $l_{s}$ are the string coupling and length respectively. The dilaton $\phi$ and the field strength $F_{4}$ of the RR 3-form $C_{3}$ are given by 
\begin{align}
e^{\phi}=g_{s}\left(\dfrac{U}{R_{\rm D4}}\right)^{3/4}\,, \qquad 
F_{4}=dC_{3}=\dfrac{2\pi N_{c}}{V_{4}}\epsilon_{4}\,,
\label{202}
\end{align}
where $V_{4}=8\pi^{2}/3$ is the volume of unit $S^{4}$ and $\epsilon_{4}$ is the corresponding volume form. In order to avoid a conical singularity at $U=U_{\rm KK}$, the $\tau$ direction should have a period of 
\begin{align}
\delta \tau 
= \dfrac{4\pi}{3}\left(\dfrac{R_{\rm D4}^{3}}{U_{\rm KK}}\right)^{1/2}
= 2\pi R = \dfrac{2\pi}{M_{\rm KK}}\,,
\label{203}
\end{align}
where $R$ is radius of $S^{1}$ and $M_{\rm KK}$ is the Kaluza--Klein mass. The parameter $U_{\rm KK}$ is related to the Kaluza--Klein mass $M_{\rm KK}$ via the relation (\ref{203}). The five dimensional gauge coupling is expressed in terms of $g_{s}$ and $l_{s}$ as $g_{\rm YM}^{2}=(2\pi)^{2}g_{s}l_{s}$. The gravity description is valid for strong coupling $\lambda \gg R$, where as usual $\lambda = g_{\rm YM}^{2}N_{c}$ denotes the 't Hooft coupling. 

The holographic QCD model at finite temperature has been proposed in \cite{ASY, PS, PSZ}. In order to introduce a finite temperature $T$ in the model, we consider the Euclidean gravitational solution which is asymptotically equals to (\ref{201}) but with the compactification of Euclidean time direction $t_{E}$. In this solution the periodicity of $t_{E}$ is arbitrary and equals to $\beta=1/T$.

Another solution with the same asymptotic is given by interchanging the role of $t_{E}$ and $\tau$ directions, 
\begin{align}
ds^{2} &=\left(\frac{U}{R_{\rm D4}}\right)^{3/2}
\Bigl(\wt{f}(U)\,(dt_{E})^{2}+(dx^{1})^{2}
+ (dx^{2})^{2}+(dx^{3})^{2} + d\tau^{2} \Bigr) \nonumber \\
& + \left(\frac{R_{\rm D4}}{U}\right)^{3/2}
\left(\frac{dU^{2}}{\wt{f}(U)}+U^{2}d\Omega_{4}^{2} \right)\,,\nonumber \\
& R_{\rm D4}^{3}=\pi g_{s}N_{c}l^{3}_{s}\,, \qquad 
\wt{f}(U)=1-\frac{U^{3}_{T}}{U^{3}}\,,
\label{204}
\end{align}
where $U_{T}$ is a parameter. To avoid a singularity at $U=U_{T}$ the period of $\delta t_{E}$ of the compactified time direction is set to 
\begin{align}
\delta t_{E} = \dfrac{4\pi}{3}\left(\dfrac{R_{\rm D4}^{3}}{U_{T}}\right)^{1/2}
=\dfrac{1}{T}\,,
\label{205}
\end{align}
and the parameter $U_{T}$ is related to the temperature $T$. The metric 
(\ref{201}) with the compactification of Euclidean time $t_{E}$ is dominant in 
the low temperature $T <1/2\pi R$, while the metric (\ref{204}) is dominant in 
the high temperature $T >1/2\pi R$. The transition between the metric 
(\ref{201}) and the metric (\ref{204}) happens when $T=1/2\pi R$. This 
transition is first-order and corresponds to the confinement/deconfinement 
phase transition in the gauge theory side. \\

\begin{center}
\scalebox{0.6}{
\unitlength 0.1in
\begin{picture}( 53.0500, 24.1500)(  1.9500,-30.3000)
%
\special{pn 8}%
\special{ar 4300 1246 800 200  0.0000000 6.2831853}%
\put(43.0000,-8.4500){\makebox(0,0){\LARGE $x_{0}$}}%
%
\special{pn 8}%
\special{ar 4300 2646 800 200  6.2831853 6.2831853}%
\special{ar 4300 2646 800 200  0.0000000 3.1415927}%
%
\special{pn 8}%
\special{pa 3500 1246}%
\special{pa 3500 2646}%
\special{fp}%
\special{pa 5100 1246}%
\special{pa 5100 2646}%
\special{fp}%
\put(13.0000,-7.0000){\makebox(0,0){\LARGE $U$}}%
%
\special{pn 8}%
\special{ar 2300 1250 800 200  0.0000000 6.2831853}%
%
\special{pn 8}%
\special{ar 2300 1250 800 1600  6.2831853 6.2831853}%
\special{ar 2300 1250 800 1600  0.0000000 3.1415927}%
%
\special{pn 13}%
\special{ar 2300 1390 600 1200  6.2831853 6.2831853}%
\special{ar 2300 1390 600 1200  0.0000000 3.1415927}%
\put(10.5000,-28.5000){\makebox(0,0){\LARGE $U_{\rm KK}$}}%
\put(11.0000,-26.0000){\makebox(0,0){\LARGE $U_{0}$}}%
\put(23.0000,-8.5000){\makebox(0,0){\LARGE $x_{4}$}}%
\put(23.0000,-22.0000){\makebox(0,0){\LARGE ${\rm D8}\text{-}\overline{\rm D8}$}}%
%
\special{pn 8}%
\special{pa 1300 3030}%
\special{pa 1300 830}%
\special{fp}%
\special{sh 1}%
\special{pa 1300 830}%
\special{pa 1280 898}%
\special{pa 1300 884}%
\special{pa 1320 898}%
\special{pa 1300 830}%
\special{fp}%
%
\special{pn 8}%
\special{pa 1300 2600}%
\special{pa 5500 2600}%
\special{dt 0.045}%
%
\special{pn 8}%
\special{ar 4300 2400 800 200  6.2831853 6.2831853}%
\special{ar 4300 2400 800 200  0.0000000 3.1415927}%
%
\special{pn 13}%
\special{pa 4970 2510}%
\special{pa 3870 1410}%
\special{dt 0.045}%
\special{pa 4920 2520}%
\special{pa 3800 1400}%
\special{dt 0.045}%
\special{pa 4880 2540}%
\special{pa 3720 1380}%
\special{dt 0.045}%
\special{pa 4830 2550}%
\special{pa 3630 1350}%
\special{dt 0.045}%
\special{pa 4780 2560}%
\special{pa 3500 1280}%
\special{dt 0.045}%
\special{pa 4730 2570}%
\special{pa 3500 1340}%
\special{dt 0.045}%
\special{pa 4670 2570}%
\special{pa 3500 1400}%
\special{dt 0.045}%
\special{pa 4620 2580}%
\special{pa 3500 1460}%
\special{dt 0.045}%
\special{pa 4570 2590}%
\special{pa 3500 1520}%
\special{dt 0.045}%
\special{pa 4510 2590}%
\special{pa 3500 1580}%
\special{dt 0.045}%
\special{pa 4450 2590}%
\special{pa 3500 1640}%
\special{dt 0.045}%
\special{pa 4390 2590}%
\special{pa 3500 1700}%
\special{dt 0.045}%
\special{pa 4330 2590}%
\special{pa 3500 1760}%
\special{dt 0.045}%
\special{pa 4270 2590}%
\special{pa 3500 1820}%
\special{dt 0.045}%
\special{pa 4210 2590}%
\special{pa 3500 1880}%
\special{dt 0.045}%
\special{pa 4150 2590}%
\special{pa 3500 1940}%
\special{dt 0.045}%
\special{pa 4090 2590}%
\special{pa 3500 2000}%
\special{dt 0.045}%
\special{pa 4030 2590}%
\special{pa 3500 2060}%
\special{dt 0.045}%
\special{pa 3960 2580}%
\special{pa 3500 2120}%
\special{dt 0.045}%
\special{pa 3890 2570}%
\special{pa 3500 2180}%
\special{dt 0.045}%
\special{pa 3820 2560}%
\special{pa 3500 2240}%
\special{dt 0.045}%
\special{pa 3740 2540}%
\special{pa 3500 2300}%
\special{dt 0.045}%
\special{pa 3660 2520}%
\special{pa 3500 2360}%
\special{dt 0.045}%
\special{pa 5010 2490}%
\special{pa 3940 1420}%
\special{dt 0.045}%
\special{pa 5050 2470}%
\special{pa 4010 1430}%
\special{dt 0.045}%
\special{pa 5080 2440}%
\special{pa 4070 1430}%
\special{dt 0.045}%
\special{pa 5100 2400}%
\special{pa 4140 1440}%
\special{dt 0.045}%
\special{pa 5100 2340}%
\special{pa 4200 1440}%
\special{dt 0.045}%
\special{pa 5100 2280}%
\special{pa 4260 1440}%
\special{dt 0.045}%
\special{pa 5100 2220}%
\special{pa 4320 1440}%
\special{dt 0.045}%
%
\special{pn 13}%
\special{pa 5100 2160}%
\special{pa 4380 1440}%
\special{dt 0.045}%
\special{pa 5100 2100}%
\special{pa 4440 1440}%
\special{dt 0.045}%
\special{pa 5100 2040}%
\special{pa 4490 1430}%
\special{dt 0.045}%
\special{pa 5100 1980}%
\special{pa 4550 1430}%
\special{dt 0.045}%
\special{pa 5100 1920}%
\special{pa 4610 1430}%
\special{dt 0.045}%
\special{pa 5100 1860}%
\special{pa 4660 1420}%
\special{dt 0.045}%
\special{pa 5100 1800}%
\special{pa 4710 1410}%
\special{dt 0.045}%
\special{pa 5100 1740}%
\special{pa 4760 1400}%
\special{dt 0.045}%
\special{pa 5100 1680}%
\special{pa 4810 1390}%
\special{dt 0.045}%
\special{pa 5100 1620}%
\special{pa 4860 1380}%
\special{dt 0.045}%
\special{pa 5100 1560}%
\special{pa 4910 1370}%
\special{dt 0.045}%
\special{pa 5100 1500}%
\special{pa 4960 1360}%
\special{dt 0.045}%
\special{pa 5100 1440}%
\special{pa 5000 1340}%
\special{dt 0.045}%
\special{pa 5100 1380}%
\special{pa 5040 1320}%
\special{dt 0.045}%
\special{pa 5100 1320}%
\special{pa 5070 1290}%
\special{dt 0.045}%
%
\special{pn 8}%
\special{ar 2300 1450 800 200  0.7853982 2.3561945}%
%
\special{pn 8}%
\special{pa 1752 1596}%
\special{pa 1734 1592}%
\special{fp}%
\special{sh 1}%
\special{pa 1734 1592}%
\special{pa 1794 1628}%
\special{pa 1786 1606}%
\special{pa 1804 1590}%
\special{pa 1734 1592}%
\special{fp}%
%
\special{pn 8}%
\special{pa 2848 1596}%
\special{pa 2866 1592}%
\special{fp}%
\special{sh 1}%
\special{pa 2866 1592}%
\special{pa 2796 1590}%
\special{pa 2816 1606}%
\special{pa 2808 1628}%
\special{pa 2866 1592}%
\special{fp}%
\put(23.0000,-18.0000){\makebox(0,0){\LARGE $L$}}%
%
\special{pn 8}%
\special{pa 2900 1250}%
\special{pa 3100 1250}%
\special{fp}%
\special{sh 1}%
\special{pa 3100 1250}%
\special{pa 3034 1230}%
\special{pa 3048 1250}%
\special{pa 3034 1270}%
\special{pa 3100 1250}%
\special{fp}%
\special{pa 2500 1250}%
\special{pa 2300 1250}%
\special{fp}%
\special{sh 1}%
\special{pa 2300 1250}%
\special{pa 2368 1270}%
\special{pa 2354 1250}%
\special{pa 2368 1230}%
\special{pa 2300 1250}%
\special{fp}%
\put(27.0000,-12.5000){\makebox(0,0){\LARGE $R$}}%
%
\special{pn 8}%
\special{pa 4900 1250}%
\special{pa 5100 1250}%
\special{fp}%
\special{sh 1}%
\special{pa 5100 1250}%
\special{pa 5034 1230}%
\special{pa 5048 1250}%
\special{pa 5034 1270}%
\special{pa 5100 1250}%
\special{fp}%
\special{pa 4500 1250}%
\special{pa 4300 1250}%
\special{fp}%
\special{sh 1}%
\special{pa 4300 1250}%
\special{pa 4368 1270}%
\special{pa 4354 1250}%
\special{pa 4368 1230}%
\special{pa 4300 1250}%
\special{fp}%
\put(47.0000,-12.5000){\makebox(0,0){\LARGE $R'$}}%
%
\special{pn 8}%
\special{pa 1300 2850}%
\special{pa 5500 2850}%
\special{dt 0.045}%
\end{picture}%
} \\
{\bf Fig. 1}\; The $\text{D8-}\overline{\text{D8}}$-branes configurations at low temperature. \\[10mm]
\end{center}

\begin{center}
\begin{tabular}{cc}
\scalebox{0.6}{
\unitlength 0.1in
\begin{picture}( 52.1000, 24.1500)(  2.9000,-30.3000)
%
\special{pn 8}%
\special{ar 4300 1246 800 200  0.0000000 6.2831853}%
\put(43.0000,-8.4500){\makebox(0,0){\LARGE $x_{0}$}}%
%
\special{pn 8}%
\special{ar 2300 2650 800 200  6.2831853 6.2831853}%
\special{ar 2300 2650 800 200  0.0000000 3.1415927}%
%
\special{pn 8}%
\special{pa 1500 1250}%
\special{pa 1500 2650}%
\special{fp}%
\special{pa 3100 1250}%
\special{pa 3100 2650}%
\special{fp}%
\put(13.0000,-7.0000){\makebox(0,0){\LARGE $U$}}%
%
\special{pn 8}%
\special{ar 2300 1250 800 200  0.0000000 6.2831853}%
%
\special{pn 8}%
\special{ar 4300 1250 800 1600  6.2831853 6.2831853}%
\special{ar 4300 1250 800 1600  0.0000000 3.1415927}%
%
\special{pn 13}%
\special{ar 2300 1390 600 1200  6.2831853 6.2831853}%
\special{ar 2300 1390 600 1200  0.0000000 3.1415927}%
\put(11.0000,-28.6000){\makebox(0,0){\LARGE $U_{\rm T}$}}%
\put(11.0000,-25.9000){\makebox(0,0){\LARGE $U_{0}$}}%
\put(23.0000,-8.5000){\makebox(0,0){\LARGE $x_{4}$}}%
\put(23.0000,-22.0000){\makebox(0,0){\LARGE ${\rm D8}\text{-}\overline{\rm D8}$}}%
%
\special{pn 8}%
\special{pa 1300 3030}%
\special{pa 1300 830}%
\special{fp}%
\special{sh 1}%
\special{pa 1300 830}%
\special{pa 1280 898}%
\special{pa 1300 884}%
\special{pa 1320 898}%
\special{pa 1300 830}%
\special{fp}%
%
\special{pn 8}%
\special{pa 1300 2590}%
\special{pa 5500 2590}%
\special{dt 0.045}%
%
\special{pn 8}%
\special{ar 2300 1450 800 200  0.7853982 2.3561945}%
%
\special{pn 8}%
\special{pa 1752 1596}%
\special{pa 1734 1592}%
\special{fp}%
\special{sh 1}%
\special{pa 1734 1592}%
\special{pa 1794 1628}%
\special{pa 1786 1606}%
\special{pa 1804 1590}%
\special{pa 1734 1592}%
\special{fp}%
%
\special{pn 8}%
\special{pa 2848 1596}%
\special{pa 2866 1592}%
\special{fp}%
\special{sh 1}%
\special{pa 2866 1592}%
\special{pa 2796 1590}%
\special{pa 2816 1606}%
\special{pa 2808 1628}%
\special{pa 2866 1592}%
\special{fp}%
\put(23.0000,-18.0000){\makebox(0,0){\LARGE $L$}}%
%
\special{pn 8}%
\special{pa 2900 1250}%
\special{pa 3100 1250}%
\special{fp}%
\special{sh 1}%
\special{pa 3100 1250}%
\special{pa 3034 1230}%
\special{pa 3048 1250}%
\special{pa 3034 1270}%
\special{pa 3100 1250}%
\special{fp}%
\special{pa 2500 1250}%
\special{pa 2300 1250}%
\special{fp}%
\special{sh 1}%
\special{pa 2300 1250}%
\special{pa 2368 1270}%
\special{pa 2354 1250}%
\special{pa 2368 1230}%
\special{pa 2300 1250}%
\special{fp}%
\put(27.0000,-12.5000){\makebox(0,0){\LARGE $R$}}%
%
\special{pn 8}%
\special{pa 4900 1250}%
\special{pa 5100 1250}%
\special{fp}%
\special{sh 1}%
\special{pa 5100 1250}%
\special{pa 5034 1230}%
\special{pa 5048 1250}%
\special{pa 5034 1270}%
\special{pa 5100 1250}%
\special{fp}%
\special{pa 4500 1250}%
\special{pa 4300 1250}%
\special{fp}%
\special{sh 1}%
\special{pa 4300 1250}%
\special{pa 4368 1270}%
\special{pa 4354 1250}%
\special{pa 4368 1230}%
\special{pa 4300 1250}%
\special{fp}%
\put(47.0000,-12.5000){\makebox(0,0){\LARGE $R'$}}%
%
\special{pn 8}%
\special{ar 4300 2390 800 200  0.9600704 2.1815223}%
%
\special{pn 13}%
\special{pa 4860 2400}%
\special{pa 3870 1410}%
\special{dt 0.045}%
\special{pa 4830 2430}%
\special{pa 3800 1400}%
\special{dt 0.045}%
\special{pa 4810 2470}%
\special{pa 3720 1380}%
\special{dt 0.045}%
\special{pa 4790 2510}%
\special{pa 3630 1350}%
\special{dt 0.045}%
\special{pa 4760 2540}%
\special{pa 3500 1280}%
\special{dt 0.045}%
\special{pa 4720 2560}%
\special{pa 3510 1350}%
\special{dt 0.045}%
\special{pa 4670 2570}%
\special{pa 3510 1410}%
\special{dt 0.045}%
\special{pa 4610 2570}%
\special{pa 3510 1470}%
\special{dt 0.045}%
\special{pa 4560 2580}%
\special{pa 3510 1530}%
\special{dt 0.045}%
\special{pa 4500 2580}%
\special{pa 3520 1600}%
\special{dt 0.045}%
\special{pa 4440 2580}%
\special{pa 3530 1670}%
\special{dt 0.045}%
\special{pa 4390 2590}%
\special{pa 3540 1740}%
\special{dt 0.045}%
\special{pa 4330 2590}%
\special{pa 3550 1810}%
\special{dt 0.045}%
\special{pa 4270 2590}%
\special{pa 3570 1890}%
\special{dt 0.045}%
\special{pa 4210 2590}%
\special{pa 3590 1970}%
\special{dt 0.045}%
\special{pa 4140 2580}%
\special{pa 3610 2050}%
\special{dt 0.045}%
\special{pa 4080 2580}%
\special{pa 3640 2140}%
\special{dt 0.045}%
\special{pa 4020 2580}%
\special{pa 3670 2230}%
\special{dt 0.045}%
\special{pa 3950 2570}%
\special{pa 3720 2340}%
\special{dt 0.045}%
\special{pa 3880 2560}%
\special{pa 3780 2460}%
\special{dt 0.045}%
\special{pa 4880 2360}%
\special{pa 3940 1420}%
\special{dt 0.045}%
\special{pa 4890 2310}%
\special{pa 4010 1430}%
\special{dt 0.045}%
\special{pa 4910 2270}%
\special{pa 4070 1430}%
\special{dt 0.045}%
\special{pa 4930 2230}%
\special{pa 4140 1440}%
\special{dt 0.045}%
\special{pa 4950 2190}%
\special{pa 4200 1440}%
\special{dt 0.045}%
\special{pa 4960 2140}%
\special{pa 4260 1440}%
\special{dt 0.045}%
\special{pa 4980 2100}%
\special{pa 4320 1440}%
\special{dt 0.045}%
\special{pa 4990 2050}%
\special{pa 4380 1440}%
\special{dt 0.045}%
\special{pa 5000 2000}%
\special{pa 4440 1440}%
\special{dt 0.045}%
\special{pa 5010 1950}%
\special{pa 4490 1430}%
\special{dt 0.045}%
%
\special{pn 13}%
\special{pa 5030 1910}%
\special{pa 4550 1430}%
\special{dt 0.045}%
\special{pa 5040 1860}%
\special{pa 4610 1430}%
\special{dt 0.045}%
\special{pa 5050 1810}%
\special{pa 4660 1420}%
\special{dt 0.045}%
\special{pa 5050 1750}%
\special{pa 4710 1410}%
\special{dt 0.045}%
\special{pa 5060 1700}%
\special{pa 4760 1400}%
\special{dt 0.045}%
\special{pa 5070 1650}%
\special{pa 4810 1390}%
\special{dt 0.045}%
\special{pa 5080 1600}%
\special{pa 4860 1380}%
\special{dt 0.045}%
\special{pa 5080 1540}%
\special{pa 4910 1370}%
\special{dt 0.045}%
\special{pa 5090 1490}%
\special{pa 4960 1360}%
\special{dt 0.045}%
\special{pa 5090 1430}%
\special{pa 5000 1340}%
\special{dt 0.045}%
\special{pa 5090 1370}%
\special{pa 5040 1320}%
\special{dt 0.045}%
\special{pa 5100 1320}%
\special{pa 5070 1290}%
\special{dt 0.045}%
%
\special{pn 8}%
\special{pa 1300 2860}%
\special{pa 5500 2860}%
\special{dt 0.045}%
\end{picture}%
} & \scalebox{0.6}{
\unitlength 0.1in
\begin{picture}( 52.3000, 24.1500)(  2.7000,-30.3000)
%
\special{pn 8}%
\special{ar 4300 1246 800 200  0.0000000 6.2831853}%
\put(43.0000,-8.4500){\makebox(0,0){\LARGE $x_{0}$}}%
\put(13.0000,-7.0000){\makebox(0,0){\LARGE $U$}}%
%
\special{pn 8}%
\special{ar 2300 1250 800 200  0.0000000 6.2831853}%
%
\special{pn 8}%
\special{ar 4300 1250 800 1600  6.2831853 6.2831853}%
\special{ar 4300 1250 800 1600  0.0000000 3.1415927}%
\put(10.8000,-28.5000){\makebox(0,0){\LARGE $U_{\rm T}$}}%
\put(23.0000,-8.5000){\makebox(0,0){\LARGE $x_{4}$}}%
\put(23.0000,-22.0000){\makebox(0,0){\LARGE ${\rm D8}\text{-}\overline{\rm D8}$}}%
%
\special{pn 8}%
\special{pa 1300 3030}%
\special{pa 1300 830}%
\special{fp}%
\special{sh 1}%
\special{pa 1300 830}%
\special{pa 1280 898}%
\special{pa 1300 884}%
\special{pa 1320 898}%
\special{pa 1300 830}%
\special{fp}%
%
\special{pn 8}%
\special{pa 1300 2850}%
\special{pa 5500 2850}%
\special{dt 0.045}%
%
\special{pn 8}%
\special{ar 2300 1450 800 200  0.7853982 2.3561945}%
%
\special{pn 8}%
\special{pa 1752 1596}%
\special{pa 1734 1592}%
\special{fp}%
\special{sh 1}%
\special{pa 1734 1592}%
\special{pa 1794 1628}%
\special{pa 1786 1606}%
\special{pa 1804 1590}%
\special{pa 1734 1592}%
\special{fp}%
%
\special{pn 8}%
\special{pa 2848 1596}%
\special{pa 2866 1592}%
\special{fp}%
\special{sh 1}%
\special{pa 2866 1592}%
\special{pa 2796 1590}%
\special{pa 2816 1606}%
\special{pa 2808 1628}%
\special{pa 2866 1592}%
\special{fp}%
\put(23.0000,-18.0000){\makebox(0,0){\LARGE $L$}}%
%
\special{pn 8}%
\special{pa 2900 1250}%
\special{pa 3100 1250}%
\special{fp}%
\special{sh 1}%
\special{pa 3100 1250}%
\special{pa 3034 1230}%
\special{pa 3048 1250}%
\special{pa 3034 1270}%
\special{pa 3100 1250}%
\special{fp}%
\special{pa 2500 1250}%
\special{pa 2300 1250}%
\special{fp}%
\special{sh 1}%
\special{pa 2300 1250}%
\special{pa 2368 1270}%
\special{pa 2354 1250}%
\special{pa 2368 1230}%
\special{pa 2300 1250}%
\special{fp}%
\put(27.0000,-12.5000){\makebox(0,0){\LARGE $R$}}%
%
\special{pn 8}%
\special{pa 4900 1250}%
\special{pa 5100 1250}%
\special{fp}%
\special{sh 1}%
\special{pa 5100 1250}%
\special{pa 5034 1230}%
\special{pa 5048 1250}%
\special{pa 5034 1270}%
\special{pa 5100 1250}%
\special{fp}%
\special{pa 4500 1250}%
\special{pa 4300 1250}%
\special{fp}%
\special{sh 1}%
\special{pa 4300 1250}%
\special{pa 4368 1270}%
\special{pa 4354 1250}%
\special{pa 4368 1230}%
\special{pa 4300 1250}%
\special{fp}%
\put(47.0000,-12.5000){\makebox(0,0){\LARGE $R'$}}%
%
\special{pn 13}%
\special{pa 1720 1390}%
\special{pa 1720 2790}%
\special{fp}%
%
\special{pn 13}%
\special{pa 2900 1380}%
\special{pa 2900 2780}%
\special{fp}%
%
\special{pn 8}%
\special{ar 2300 2650 800 200  6.2831853 6.2831853}%
\special{ar 2300 2650 800 200  0.0000000 3.1415927}%
%
\special{pn 8}%
\special{pa 1510 1250}%
\special{pa 1510 2650}%
\special{fp}%
\special{pa 3110 1250}%
\special{pa 3110 2650}%
\special{fp}%
%
\special{pn 13}%
\special{pa 4690 2650}%
\special{pa 3510 1470}%
\special{dt 0.045}%
\special{pa 4710 2610}%
\special{pa 3510 1410}%
\special{dt 0.045}%
\special{pa 4740 2580}%
\special{pa 3510 1350}%
\special{dt 0.045}%
\special{pa 4770 2550}%
\special{pa 3500 1280}%
\special{dt 0.045}%
\special{pa 4790 2510}%
\special{pa 3630 1350}%
\special{dt 0.045}%
\special{pa 4810 2470}%
\special{pa 3720 1380}%
\special{dt 0.045}%
\special{pa 4830 2430}%
\special{pa 3800 1400}%
\special{dt 0.045}%
\special{pa 4860 2400}%
\special{pa 3870 1410}%
\special{dt 0.045}%
\special{pa 4880 2360}%
\special{pa 3940 1420}%
\special{dt 0.045}%
\special{pa 4890 2310}%
\special{pa 4010 1430}%
\special{dt 0.045}%
\special{pa 4910 2270}%
\special{pa 4070 1430}%
\special{dt 0.045}%
\special{pa 4930 2230}%
\special{pa 4140 1440}%
\special{dt 0.045}%
\special{pa 4950 2190}%
\special{pa 4200 1440}%
\special{dt 0.045}%
\special{pa 4960 2140}%
\special{pa 4260 1440}%
\special{dt 0.045}%
\special{pa 4980 2100}%
\special{pa 4320 1440}%
\special{dt 0.045}%
\special{pa 4990 2050}%
\special{pa 4380 1440}%
\special{dt 0.045}%
\special{pa 5000 2000}%
\special{pa 4440 1440}%
\special{dt 0.045}%
\special{pa 5010 1950}%
\special{pa 4490 1430}%
\special{dt 0.045}%
\special{pa 5030 1910}%
\special{pa 4550 1430}%
\special{dt 0.045}%
\special{pa 5040 1860}%
\special{pa 4610 1430}%
\special{dt 0.045}%
\special{pa 5050 1810}%
\special{pa 4660 1420}%
\special{dt 0.045}%
\special{pa 5050 1750}%
\special{pa 4710 1410}%
\special{dt 0.045}%
\special{pa 5060 1700}%
\special{pa 4760 1400}%
\special{dt 0.045}%
\special{pa 5070 1650}%
\special{pa 4810 1390}%
\special{dt 0.045}%
\special{pa 5080 1600}%
\special{pa 4860 1380}%
\special{dt 0.045}%
\special{pa 5080 1540}%
\special{pa 4910 1370}%
\special{dt 0.045}%
\special{pa 5090 1490}%
\special{pa 4960 1360}%
\special{dt 0.045}%
\special{pa 5090 1430}%
\special{pa 5000 1340}%
\special{dt 0.045}%
\special{pa 5090 1370}%
\special{pa 5040 1320}%
\special{dt 0.045}%
\special{pa 5100 1320}%
\special{pa 5070 1290}%
\special{dt 0.045}%
%
\special{pn 13}%
\special{pa 4660 2680}%
\special{pa 3510 1530}%
\special{dt 0.045}%
\special{pa 4630 2710}%
\special{pa 3520 1600}%
\special{dt 0.045}%
\special{pa 4590 2730}%
\special{pa 3530 1670}%
\special{dt 0.045}%
\special{pa 4560 2760}%
\special{pa 3540 1740}%
\special{dt 0.045}%
\special{pa 4520 2780}%
\special{pa 3550 1810}%
\special{dt 0.045}%
\special{pa 4480 2800}%
\special{pa 3570 1890}%
\special{dt 0.045}%
\special{pa 4440 2820}%
\special{pa 3590 1970}%
\special{dt 0.045}%
\special{pa 4400 2840}%
\special{pa 3610 2050}%
\special{dt 0.045}%
\special{pa 4350 2850}%
\special{pa 3640 2140}%
\special{dt 0.045}%
\special{pa 4290 2850}%
\special{pa 3670 2230}%
\special{dt 0.045}%
\special{pa 4220 2840}%
\special{pa 3720 2340}%
\special{dt 0.045}%
\special{pa 4130 2810}%
\special{pa 3780 2460}%
\special{dt 0.045}%
\end{picture}%
} \\
\end{tabular}
{\bf Fig. 2}\; The $\text{D8-}\overline{\text{D8}}$-branes configurations at 
high temperature. \\[10mm]
\end{center}

In the low temperature, the $\text{D8-}$ and $\overline{\text{D8}}$-branes are 
connected at $U=U_{0}$ as shown in Fig. 1. The connected 
configuration of the $\text{D8-}\overline{{\text{D8}}}$-branes indicates that 
the ${\rm U}({\rm N}_{f})_{L} \times {\rm U}({\rm N}_{f})_{R}$ global chiral symmetry is broken to a diagonal subgroup ${\rm U}({\rm N}_{f})$. We refer to the  connected configuration in the low temperature as the {\it low-temperature 
phase}. In the high temperature, there are two kinds of configurations as 
shown in Fig. 2. One is connected configuration and the other is disconnected 
configuration that the $\text{D8-}$ and $\overline{\text{D8}}$-branes hang 
vertically from infinity down to the horizon. The disconnected configuration 
of the $\text{D8-}\overline{{\text{D8}}}$-branes indicates that the 
${\rm U}({\rm N}_{f})_{L} \times {\rm U}({\rm N}_{f})_{R}$ global chiral 
symmetry is restored. We refer to the disconnected configuration and the 
connected configuration in the high temperature as the 
{\it high-temperature phase} and the {\it intermediate-temperature phase}, 
respectively. The intermediate-temperature phase is realized when the 
confinement/deconfinement phase transition and the chiral phase transition does 
not occur simultaneously. 
If the ratio $L/R$ is larger than $0.97$, there is no intermediate-temperature phase \cite{ASY}.

When turning on a NS-NS $B$-field on the D-brane worldvolume, the low-energy 
effective worldvolume theories are deformed to a noncommutative Yang-Mills 
theories \cite{CDS, DH, AASJ, SJ, SW}. The D-brane realizations of 
noncommutative Yang-Mills theories have a gravity dual in the large $N$, 
strong 't Hooft coupling limit \cite{HI, MR, AOSJ}. 
In accordance with the formulation of 
\cite{HI, MR, AOSJ}, we attempt to construct the gravity dual of the noncommutative QCD whose chiral symmetry is spontaneously broken by deforming the 
holographic QCD model. Let us consider the D4-branes solution compactified on a circle in the $\tau$-direction. T-dualizing it along $x^{3}$ produces a 
D3-branes delocalized along $x^{3}$. After rotating the D3-branes along the 
($x^{2},\;x^{3}$) plane, we T-dualize back on $x^{3}$. This procedure yields 
the solution with a $B_{23}$ fields along the $x^{2}$ and $x^{3}$ directions. 
The solution in the low temperature takes the form 
\begin{align}
ds^{2} &=\left(\frac{U}{R_{\rm D4}}\right)^{3/2}
\Bigl((dt_{E})^{2}+(dx^{1})^{2}
+ h\{(dx^{2})^{2}+(dx^{3})^{2}\} + f(U)\,d\tau^{2} \Bigr) 
\nonumber \\
&+ \left(\frac{R_{\rm D4}}{U}\right)^{3/2}
\left(\frac{dU^{2}}{f(U)}+U^{2}d\Omega_{4}^{2} \right)\,,
\label{206}
\end{align}
where $h(U)=\dfrac{1}{1+\theta^{3}U^{3}}$ and $\theta$ denotes the 
noncommutativity parameter with dimension of mass. When $\theta \neq 0$ this 
solution is dual to a gauge theory in which the coordinates $x^{2}$ and $x^{3}$ do not commute. It is obvious that this solution reduces to the solution 
(\ref{201}) with Euclidean signature when $\theta =0$. In the high 
temperature, the solution (\ref{206}) changes to 
\begin{align}
ds^{2} &=\left(\frac{U}{R_{\rm D4}}\right)^{3/2}
\Bigl(\wt{f}(U)(dx_{E})^{2}+(dx^{1})^{2}
+ h\{(dx^{2})^{2}+(dx^{3})^{2}\} + d\tau^{2} \Bigr) 
\nonumber \\
&+ \left(\frac{R_{\rm D4}}{U}\right)^{3/2}
\left(\frac{dU^{2}}{\wt{f}(U)}+U^{2}d\Omega_{4}^{2} \right)\,. 
\label{207}
\end{align}
The solution has the same form as the one in the low temperature (\ref{206}), but with the role of the $\tau$ and $t_{E}$ directions exchanged.

The effective action of the D8 branes is given as the Dirac-Born-Infeld(DBI) action with Wess-Zumino-like(WZ) term,
\begin{align}
\label{208}
S_{\rm D8} &=S^{\rm DBI}_{\rm D8} + S^{\rm WZ}_{\rm D8}\,, \\ 
& S^{\rm DBI}_{\rm D8}
=T_{8}\int d^{9}x e^{-\phi}{\rm Tr}\sqrt{\det (g_{MN}+B_{MN} 
+ 2\pi l_{s}{}^{2}F_{MN})}\,, 
\nonumber \\
& S^{\rm WZ}_{\rm D8}
=T_{8}\int_{\rm D8}\,C_{3} \wedge e^{(\,B+2\pi l_{s}{}^{2}F\,)}\,, 
\nonumber 
\end{align}
where $T_{8}=1/(2\pi)^{8}l_{s}^{9}$ is the tension of the D8 brane, $g_{MN}$ and $F_{MN}\;(M,N=0,1,\ldots,8)$ are the induced metric and the field strength of the $U(N_{f})$ gauge field $A_{M}$ on the D8-brane, respectively. 
We regard the gauge field on the D8-brane as a background gauge field. For 
simplicity, we assume that $A_{0}$ and $A_{1}$ components of the $U(1)$ part of the gauge field are non-vanishing fields. As will be mentioned later, the 
boundary value of $A_{0}$ is related to the baryon number chemical potential. 
The $B$-field is not a constant field which promotes chiral symmetry breaking 
in the QCD \cite{BLL, JK, H}. The notation $B_{MN}$ denotes the pullback of the NS-NS $B$-field\,:
\begin{align}
B_{MN}(U)=\left\{
\begin{array}{ll}
\theta^{3/2}\dfrac{U^{3}}{R_{\rm D4}^{3/2}}h(U) & (M=2,\;N=3) \\
0 & (\text{others})
\end{array} \right.\;.
\label{209}
\end{align}
We make an ansatz that $A_{0}$ and $\tau$ depend only on the coordinate $U$\,: 
\begin{align}
\tau=\tau(U)\,,\quad A_{0}=A_{0}(U)\,, 
\label{210}
\end{align}
and $A_{1}$ is not an auxiliary field but a non-zero constant. 

%
%

\section{Chiral phase restoration in the noncommutative QCD} 

\setcounter{equation}{0}
\addtocounter{enumi}{1}

\subsection{Low temperature} 

In order to determine the chiral phase of the noncommutative QCD, we need to 
analyze the shape of the probe D8-branes. The induced metric on the probe 
D8-branes associated with the confining phase reads
\begin{align}
ds^{2} &=\left(\frac{U}{R_{\rm D4}}\right)^{3/2}
\Bigl((dt_{E})^{2}+(dx^{1})^{2} + h\{(dx^{2})^{2}+(dx^{3})^{2}\} \Bigr) 
+ \left(\frac{R_{\rm D4}}{U}\right)^{3/2}U^{2}d\Omega_{4}^{2} \nonumber \\
& + \left[ \left(\frac{U}{R_{\rm D4}}\right)^{3/2} f(U)(\tau'(U))^{2}
+ \left(\frac{R_{\rm D4}}{U}\right)^{3/2} \frac{1}{f(U)} \right] dU^{2} \;,
\label{301}
\end{align}
where $\tau'=\dfrac{d\tau}{dU}$. Substituting the determinant of the induced 
metric (\ref{301}) and the dilaton in the $B$-field background $e^{\phi}=g_{s}h^{1/2}\left(\dfrac{U}{R_{\rm D4}}\right)^{3/4}$ into the DBI action (\ref{208}), we have 

\begin{align}
S^{\rm DBI}_{\rm D8} = \dfrac{N_{f}T_{8}V_{4}}{g_{s}} 
\int d^{4}x dU\,U^{4}
\sqrt{f\tau'^{2} + \left(\dfrac{R_{\rm D4}}{U}\right)^{3}(f^{-1}
-(2\pi l_{s}{}^{2}A'_{0})^{2})} \;,
\label{302}
\end{align}
where $A_{0}'=\dfrac{dA_{0}}{dU}$. We notice that the $B$-field in the DBI 
determinant $\det (g_{MN}+B_{MN} + 2\pi l_{s}{}^{2}F_{MN})$ is dropped out and 
there is no dependence on the noncommutativity parameter $\theta$ in the DBI 
action. The independence of the DBI action on the noncommutativity parameter 
also takes place in the effective action of the probe D7-brane \cite{APR}. 
Adding the relevant WZ-term 
\begin{align}
\label{208a}
S_{\rm D8}^{\rm WZ}=T_{8}\dfrac{(2\pi l_{s})^{2}}{2}
\int_{\rm D8} C_{3} \wedge B \wedge F \wedge F \, 
\end{align}
to the DBI action and using the relation (\ref{202}), we 
find the term involving the $B$-field in the effective action of the D8-branes
\begin{align}
\label{303}
S_{\rm D8} &=S^{\rm DBI}_{\rm D8}+S^{\rm WZ}_{\rm D8} \\
&= \dfrac{N_{f}T_{8}V_{4}}{g_{s}} 
\int d^{4}x dU\,U^{4}\left[\,
\sqrt{f\tau'^{2} + \left(\dfrac{R_{\rm D4}}{U}\right)^{3}
(f^{-1}-(2\pi l_{s}{}^{2}A'_{0})^{2})}
+\kappa BU^{-4}A'_{0} \,\right] \,, \nonumber 
\end{align}
where $\kappa=\dfrac{N_{c}}{12\pi^{3}l_{s}^{2}}\dfrac{g_{s}A_{1}}{N_{f}T_{8}V_{4}}$ and $B = B_{23}$. Thus, the influence of the space noncommutativity is 
brought from WZ-term. The equation of motions for $\tau$ and $A_{0}$ are 
\begin{align}
& \dfrac{d}{dU}\Biggl[ 
\dfrac{U^{4}f\tau'}{\sqrt{f\tau'^{2}+\left(\dfrac{R_{\rm D4}}{U}\right)^{3}
\Bigl(f^{-1}-(2\pi l_{s}{}^{2}A'_{0})^{2}\Bigr)}}\Biggr]=0 \,, 
\label{304} \\
& \dfrac{d}{dU}\Biggl[ 
\dfrac{U^{4}\left(\dfrac{R_{\rm D4}}{U}\right)^{3}(2\pi l_{s}{}^{2}A'_{0})}
{\sqrt{f\tau'^{2}+\left(\dfrac{R_{\rm D4}}{U}\right)^{3}
\Bigl(f^{-1}-(2\pi l_{s}{}^{2}A'_{0})^{2}\Bigr)}}
-\dfrac{\kappa B}{2\pi l_{s}{}^{2}}\Biggr]=0 \,, 
\label{305}
\end{align}
respectively. Integrating the equations of motion (\ref{304}) and (\ref{305}), 
we obtain
\begin{align}
& (\tau')^{2}
= \dfrac{\left(\dfrac{R_{\rm D4}}{U}\right)^{6}
\left\{U_{0}^{8}+\left(C+\dfrac{\kappa B_{0}}{2\pi l_{s}{}^{2}}\right)^{2}
\left(\dfrac{U_{0}}{R_{\rm D4}}\right)^{3} \right\}f_{0}}
{f^{2}\left[\left(\dfrac{R_{\rm D4}}{U}\right)^{3}
(U^{8}f-U_{0}^{8}f_{0}) 
+ \left(C+\dfrac{\kappa B}{2\pi l_{s}{}^{2}}\right)^{2}f 
- \left(C+\dfrac{\kappa B_{0}}{2\pi l_{s}{}^{2}}\right)^{2}
\left(\dfrac{U_{0}}{U}\right)^{3}f_{0} \right]} \,, 
\label{306} \\
& (2\pi l_{s}{}^{2}A'_{0})^{2}
= \dfrac{\left(C+\dfrac{\kappa B}{2\pi l_{s}{}^{2}} \right)^{2}}
{\left(\dfrac{R_{\rm D4}}{U}\right)^{3}
(U^{8}f-U_{0}^{8}f_{0}) 
+ \left(C+\dfrac{\kappa B}{2\pi l_{s}{}^{2}}\right)^{2}f 
- \left(C+\dfrac{\kappa B_{0}}{2\pi l_{s}{}^{2}}\right)^{2}
\left(\dfrac{U_{0}}{U}\right)^{3}f_{0}} \,, 
\label{307}
\end{align}
where $f_{0}=f(U_{0})$, $B_{0}=B(U_{0})$, and $C$ denotes a constant of integration. In deriving the equations (\ref{306}) and (\ref{307}), we impose the condition $\tau'(U_{0}) \to \infty$ that corresponds to the connected configuration of the $\text{D8-}\overline{\text{D8}}$-branes. In order to avoid the singularity 
of $A'_{0}$ at $U = U_{0}$, the constant of integration in Eq. (\ref{307}) 
should be chosen to $C=-\dfrac{\kappa B_{0}}{2\pi l_{s}^{2}}$. Since the 
asymptotic behavior of $A'_{0}$ as $U \to \infty$ is $A'_{0} \sim 0$, we 
obtain 
\begin{align}
\mu = \lim_{U \to \infty}A_{0}(U) \,,
\label{308}
\end{align}
where $\mu$ denotes a finite constant. The boundary value $\mu$ coupled to the baryon number density and can be regarded as the baryon number chemical 
potential \cite{KSZ, HT, PS2}. 

Substituting (\ref{306}) and (\ref{307}) into the D8-branes action (\ref{303}), we have 
\begin{align}
\label{309}
S_{\rm D8}^{\rm U} &= \widetilde{T}_{8} 
\int^{\infty}_{0} du\,\{\,u^{5}+b\,(u)\,(b\,(u)-b\,(1))\,\}
\sqrt{\dfrac{u^{3}}
{u^{3}f(u)\{u^{5}+(b\,(u)-b\,(1))^{2}\}-f(1)}} \,, 
\end{align}
where $u = \dfrac{U}{U_{0}}$ is a dimensionless variable, $f(u)=1-\left(\dfrac{U_{\rm KK}}{U_{0}}\right)^3 u^{-3}$, and 
$\widetilde{T}_{8}=\dfrac{N_{f}T_{8}V_{4}}{g_{s}}(R_{\rm D4}^{3}U_{0}^{7})^{1/2}\displaystyle{\int}d^{4}x$. The notation ``${\rm U}$'' at the upper right of $S_{\rm D8}$ denotes the connected configuration of the $\text{D8-}$ and $\overline{\text{D8}}$-branes. The notation $b$ denotes the dimensionless $B$-field 
defined by 
\begin{align}
\label{310}
b\,(u)=\dfrac{1}{R_{\rm D4}^{3/2}U_{0}^{5/2}}
\dfrac{\kappa B}{2\pi l_{s}{}^{2}}
= 12\pi^2 \dfrac{(2\pi l_{s}{}^{2} A_{1})}{N_{f}U_{0}}
\dfrac{q^{3/2}u^{3}}{1+q^{3}u^{3}}\,,
\end{align}
where $q=U_{0}\theta$ is the dimensionless noncommutativity parameter. When 
$B$-field is absent, $b$ equals to zero and the effective action (\ref{309}) 
coincides with the result in ref. \cite{HT}.


\subsection{High temperature} 

The induced metric on the probe D8-branes associated with the deconfining phase takes the form 
\begin{align}
\label{311} 
ds^{2} &=\left(\frac{U}{R_{\rm D4}}\right)^{3/2}
\Bigl(\widetilde{f}(U)(dt_{E})^{2}+(dx^{1})^{2} 
+ h\{(dx^{2})^{2}+(dx^{3})^{2}\} \Bigr) 
+ \left(\frac{R_{\rm D4}}{U}\right)^{3/2}U^{2}d\Omega_{4}^{2} \nonumber \\
& + \left[ \left(\frac{U}{R_{\rm D4}}\right)^{3/2} (\tau'(U))^{2}
+ \left(\frac{R_{\rm D4}}{U}\right)^{3/2} \frac{1}{\widetilde{f}(U)} \right] 
dU^{2} \;.
\end{align}
Substituting the determinant of the induced metric (\ref{311}) and the dilaton (\ref{202}) into the effective action of the D8 branes (\ref{208}), we have 
\begin{align}
\label{312}
S_{\rm D8}
&=S^{\rm DBI}_{\rm D8}+S^{\rm WZ}_{\rm D8} \nonumber \\ 
&= \dfrac{N_{f}T_{8}V_{4}}{g_{s}} 
\int d^{4}x dU\,U^{4}
\left[\, \sqrt{\widetilde{f}\tau'^{2} 
+ \left(\dfrac{R_{\rm D4}}{U}\right)^{3}(1-(2\pi l_{s}{}^{2}A'_{0})^{2})}
+\kappa BU^{-4}A'_{0} \,\right]\;.
\end{align}
The dependence of the effective action of the D8 branes on the $B$-field 
arises from the WZ term in the same way as in the low-temperature phase. The 
equations of motion for $\tau$ and $A_{0}$ are 
\begin{align}
& \dfrac{d}{dU}
\Biggl[\dfrac{ U^{4}\wt{f}\tau'}{\sqrt{
\wt{f}\tau'^{2} + \left(\dfrac{R_{\rm D4}}{U}\right)^{3}
\Bigl( 1-(2\pi l_{s}{}^{2}A'_{0})^{2} \Bigr)}} \Biggr]=0\,, 
\label{313} \\
& \dfrac{d}{dU}\Biggl[\dfrac{U^{4}\left(\dfrac{R_{\rm D4}}{U}\right)^{3}
(2\pi l_{s}{}^{2}A'_{0})}{\sqrt{
\wt{f}\tau'^{2} + \left(\dfrac{R_{\rm D4}}{U}\right)^{3}
\Bigl( 1-(2\pi l_{s}{}^{2}A'_{0})^{2} \Bigr)}} 
- \dfrac{\kappa B}{2\pi l_{s}{}^{2}}
\Biggr]=0 \,, 
\label{314}
\end{align}
respectively. There is possibility of taking two kinds of the configurations 
of the $\text{D8-}$ and $\overline{\text{D8}}$-branes in the high temperature. 
We consider first the connected configuration of the $\text{D8-}
\overline{\text{D8}}$-branes. Integrating the equation of motions (\ref{313}) 
and (\ref{314}) and imposing the condition $\tau'(U) \to \infty$, we have
\begin{align}
& (\tau')^{2}
= \dfrac{U_{0}^{8}\left(\dfrac{R_{\rm D4}}{U}\right)^{6}\wt{f}_{0}}
{\wt{f}^{2}\left[\left(\dfrac{R_{\rm D4}}{U}\right)^{3}
(U^{8}\wt{f}-U_{0}^{8}\wt{f}_{0}) 
+ \left(\dfrac{\kappa}{2\pi l_{s}^{2}}\right)^{2}
(B-B_{0})^{2} \wt{f} \right]} \,, \label{315} \\
& (2\pi\alpha'A'_{0})^{2}
= \dfrac{\wt{f}\left(\dfrac{\kappa}{2\pi\alpha'} \right)^{2}(B-B_{0})^{2}}
{\left(\dfrac{R_{\rm D4}}{U}\right)^{3}
(U^{8}\wt{f}-U_{0}^{8}\wt{f}_{0}) 
+ \left(\dfrac{\kappa}{2\pi l_{s}^{2}} \right)^{2}(B-B_{0})^{2}\wt{f}} \,, 
\label{316}
\end{align}
where $\wt{f}_{0}=\wt{f}(U_{0})$. The constant of integration is chosen to avoid the singularity of $A'_{0}$ at $U=U_{0}$. Notice that $A'_{0}(U)$ becomes always zero and $A_{0}(U)$ takes a constant value when we turn off the $B$-field. 
By virtue of the presence of non-zero $B$-field, $A_{0}(U)$ is not a constant. 

Substitution of (\ref{315}) and (\ref{316}) into the effective action (\ref{312}) leads to 
\begin{align}
S^{\rm U}_{\rm D8}=\wt{T}_{8} 
\int^{\infty}_{1} du\, \{u^{5}+b\,(u)\,(b\,(u)-b\,(1))\}
\,\sqrt{\dfrac{u^{3}\widetilde{f}(u)}{u^{3}
\wt{f}(u)\{ u^{5}+(b\,(u)-b\,(1))^{2} \} 
- \wt{f}(1)}} \,. \label{317} 
\end{align}
where $u$ and $\widetilde{T}_{8}$ are the same notations as in Eq. (\ref{309}).

The disconnected configuration of the $\text{D8-}$ and $\overline{\text{D8}}$-branes corresponds to a condition $\tau'(U)=0$. For the other condition $\tau'(U)=0$, we obtain 
\begin{align}
(2\pi\alpha'A'_{0})^{2}
= \dfrac{\left(\widetilde{C} + \dfrac{\kappa B}{2\pi l_{s}^{2}}\right)^{2}}
{U^{8}\left(\dfrac{R_{\rm D4}}{U}\right)^{3}+\left(\widetilde{C} + \dfrac{\kappa B}{2\pi l_{s}^{2}}\right)^{2}} \,, \label{318} 
\end{align}
where $\wt{C}$ is a constant of integration. For both Eq (\ref{316}) and 
(\ref{318}), the asymptotic value of $A'_{0}$ for $U \to \infty$ is zero, 
\begin{align}
\lim_{U \to \infty}(2\pi\alpha'A'_{0})^{2}
= 0 \,. \label{319} 
\end{align}
From (\ref{319}) the asymptotic value of $A_{0}$ is a constant. We are able to 
choose the same asymptotic value of $A_{0}$ in both the connected and 
disconnected configuration as 
\begin{align}
\lim_{U \to \infty}A_{0}(U)=\mu \,. \label{320} 
\end{align}
The constant value $\mu$ is regarded as the chemical potential. From Eq. 
(\ref{318}) and Eq. (\ref{320}), the chemical potential is expressed in terms 
of integration with respect to $u$
\begin{align}
\mu=A_{0}(U \to \infty) 
= \dfrac{U_{0}}{2\pi l_{s}^{2}} \int^{u}_{u_{T}}
\;du'\sqrt{\dfrac{(\wt{c}+b)^{2}}{u'^{5}+(\wt{c}+b)^{2}}}\,. \label{321}
\end{align}

Substituting $\tau'(U)=0$ and (\ref{318}) into the action (\ref{312}), we have 
\begin{align}
S^{\|}_{\rm D8}
= \widetilde{T}_{8}\int^{\infty}_{u_{T}} \; du 
\dfrac{u^{5}+b\,(b+c)} {\sqrt{u^{5}+(b+c)^{2}}} \,, \label{322}
\end{align}
where $c=\dfrac{\wt{C}}{R_{\rm D4}^{3/2}U_{0}^{5/2}}$ is a dimensionless 
constant. The notation ``$\|$'' at the upper right of $S_{\rm D8}$ denotes the 
disconnected configuration of the $\text{D8-}$ and 
$\overline{\text{D8}}$-branes. The difference between the two effective 
actions (\ref{317}) and (\ref{322}) is given by \\

\begin{center}
\begin{tabular}{cc}
\scalebox{0.8}{\includegraphics[width=100mm]{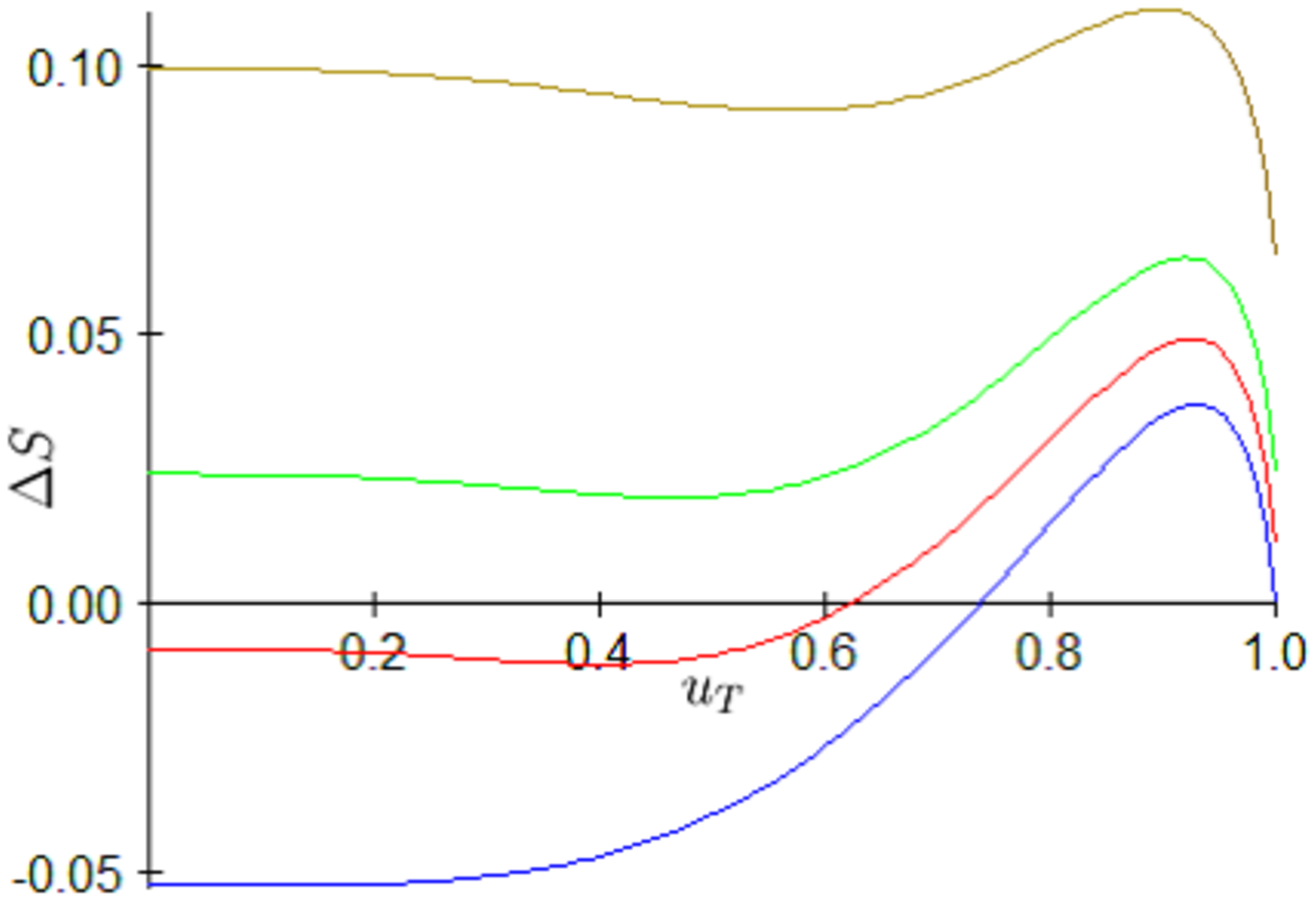}} & 
\scalebox{0.8}{\includegraphics[width=100mm]{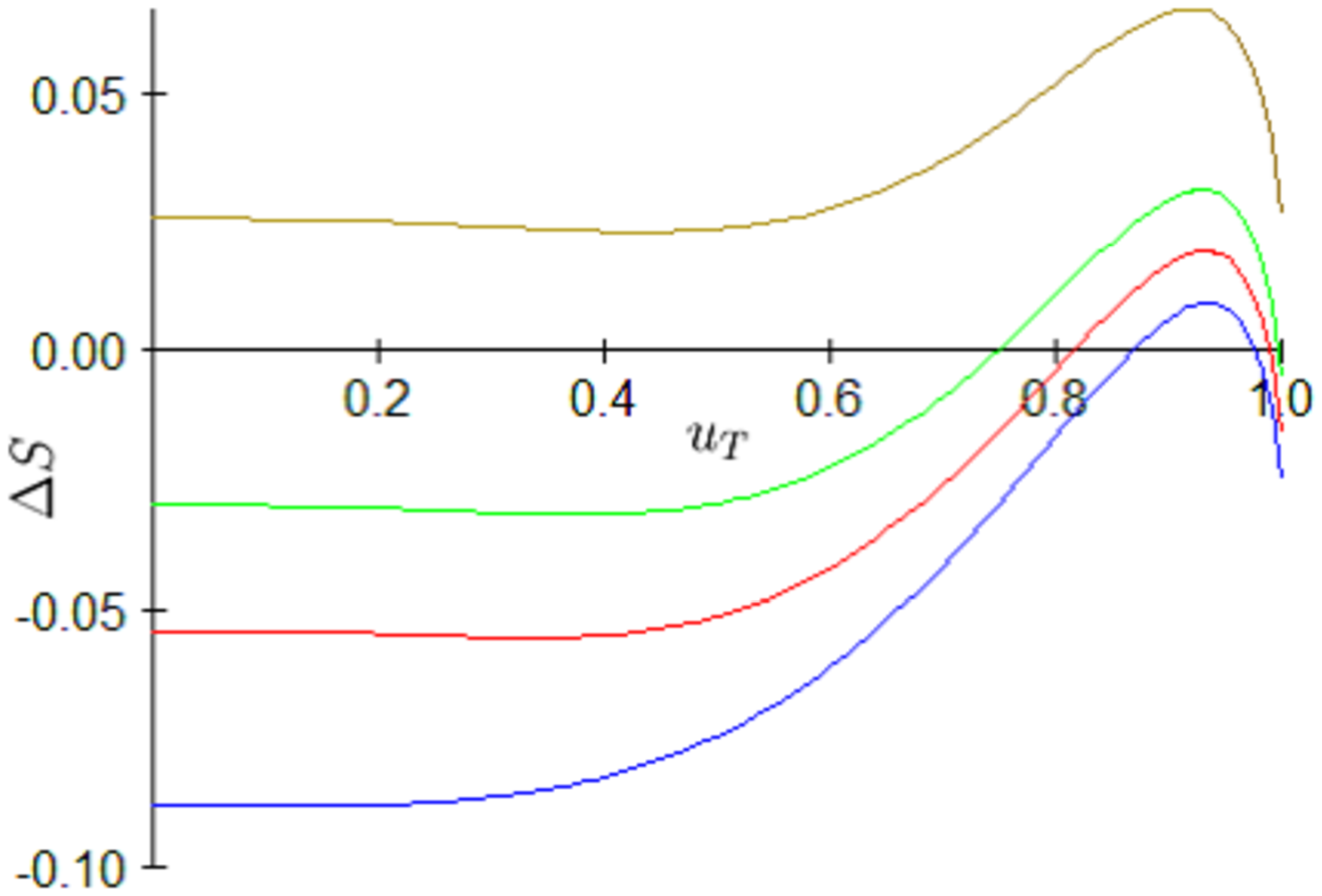}} \\
$q=0.0$ & $q=0.5$ \\
\scalebox{0.8}{\includegraphics[width=100mm]{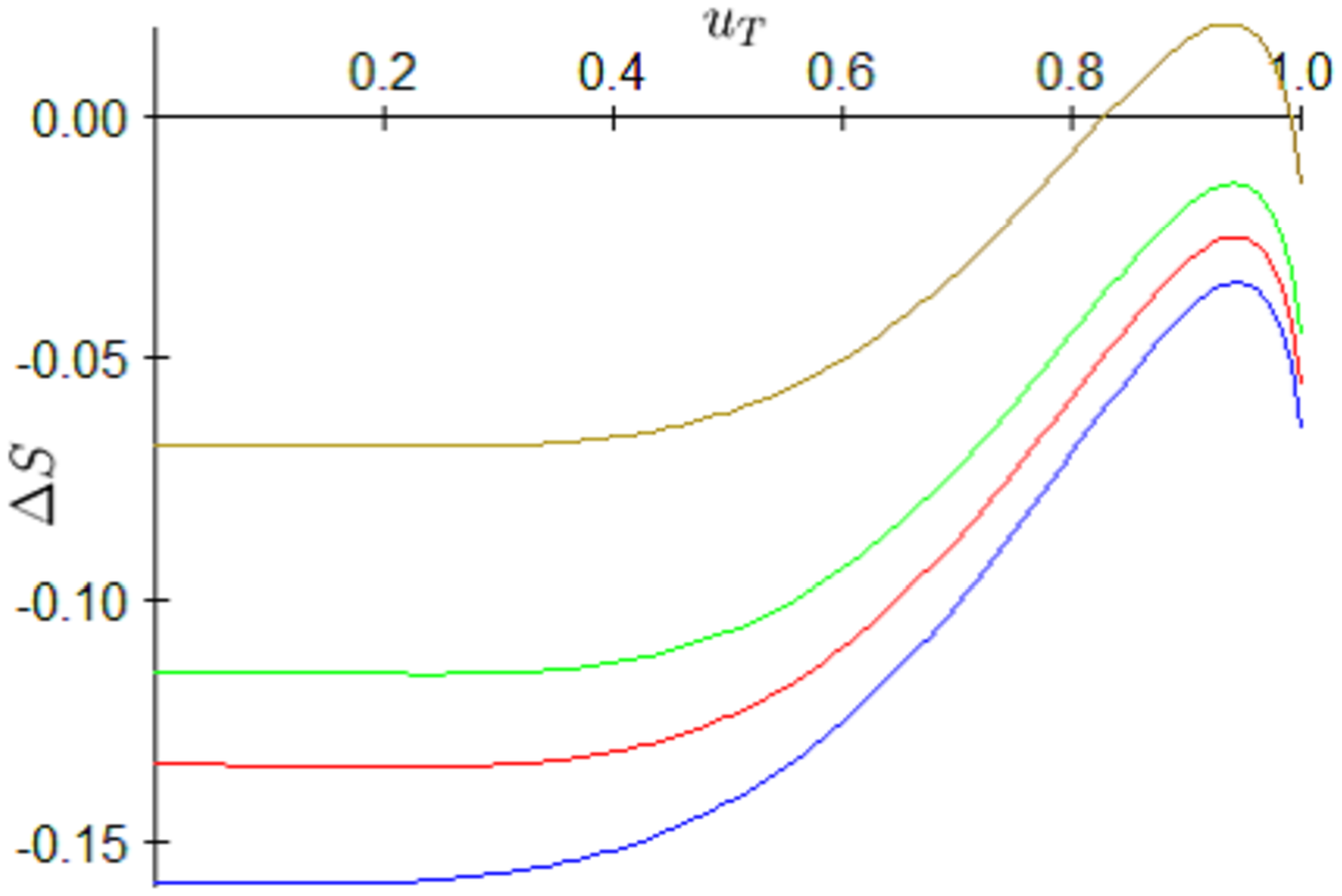}} & 
\scalebox{0.8}{\includegraphics[width=100mm]{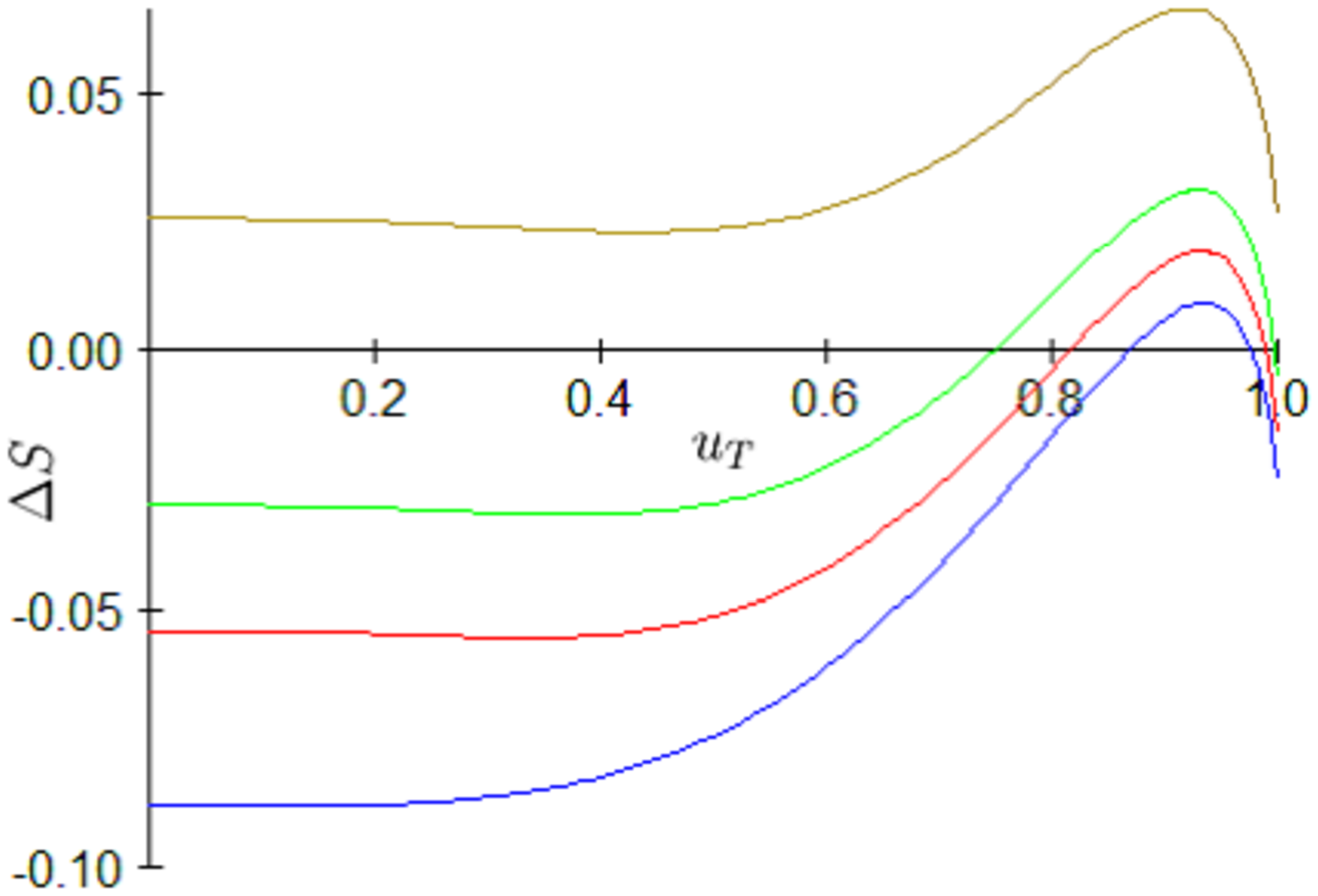}} \\
$q=1.0$ & $q=5.0$ 
\end{tabular}
\end{center}
{\bf Fig. 3}\; The numerical calculation of $\Delta S$ as a function of $u_{T}$ for some values of $c$ and $q$. From the bottom to the top each line represents the value of $c$ for $0.0,\;0.2,\;0.3,\;0.5,$ respectively. \\[10mm]

\begin{align}
\label{323}
\Delta S 
& \equiv \dfrac{S^{\rm U}_{\rm D8}-S^{\|}_{\rm D8}}{\widetilde{T}_{8}} \\ 
&= \int^{\infty}_{1} du\, \{u^{5}+b\,(b-b_{0})\}
\,\sqrt{\dfrac{u^{3}\widetilde{f}(u)}
{u^{3}\widetilde{f}(u)\{ u^{5}+(b-b_{0})^{2} \} 
- \widetilde{f}(1)}} 
-\int^{\infty}_{u_{T}} \; du 
\dfrac{u^{5}+b\,(b+c)}
{\sqrt{u^{5}+(b+c)^{2}}}\,. \nonumber
\end{align}
The sign of $\Delta S$ indicates which phase is realized, the chiral symmetry phase or the chiral symmetry broken phase. For $\Delta S <0$, the chiral symmetry broken phase is dominant, while for $\Delta S >0$, the chiral symmetry phase is dominant. The result of the numerical calculation of Eq. (\ref{323}) is shown 
in Fig. 3.

For $q=0$, the difference $\Delta S$ is reduced to the commutative theory and 
the behavior of $\Delta S$ is in agreement with the result in Ref. \cite{HT}. 
As the parameter $c$ increases, the phase transition point $u_{T}$ where 
$\Delta S=0$ decreases for a fixed $q$. In commutative theory, there is a 
value of $c$ where $\Delta S$ is positive for all values of $u_{T}$ and the 
chiral symmetry is always restored. In contrast, the phase transition point 
$u_{T}$ increases with increasing $q$ for a fixed parameter $c$. The 
noncommutativity parameter $q$ plays the similar role of the constant external 
$B$-field \cite{JK}. In noncommutative theory, there is a value of $c$ where 
$\Delta S$ is negative for all values of $u_{T}$ and the chiral symmetry is 
always broken. However, the Fig. 3 shows that the difference $\Delta S$ for 
$q \to \infty$ reduces to the one in the commutative theory. This fact is 
known as a characteristic of the noncommutative theory \cite{APR}.

Fig. 3 indicates that the chiral symmetry phase structure of this model depends 
on the parameter $u_{T},\;c$ and the dimensionless noncommutativity parameter 
$q$. This fact allowed us to draw a phase diagram for the noncommutative QCD 
in the $(c,\;u_{T})$ plane. The phase diagrams with $q=0.0$ and $q=1.0$ are 
shown in Fig. 4. \\

\begin{center}
\begin{tabular}{cc}
\scalebox{0.8}{\includegraphics[width=100mm]{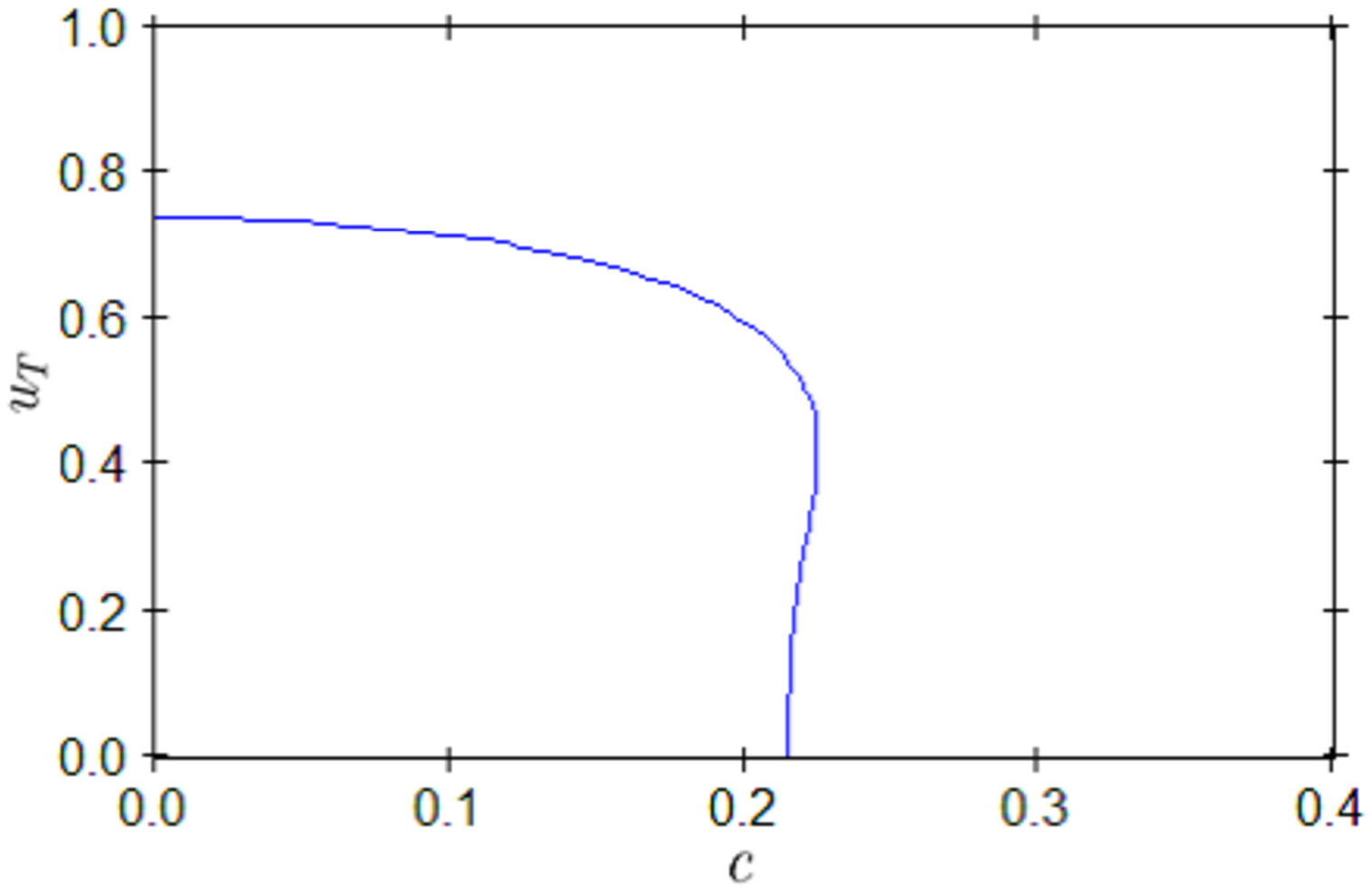}} & 
\scalebox{0.8}{\includegraphics[width=100mm]{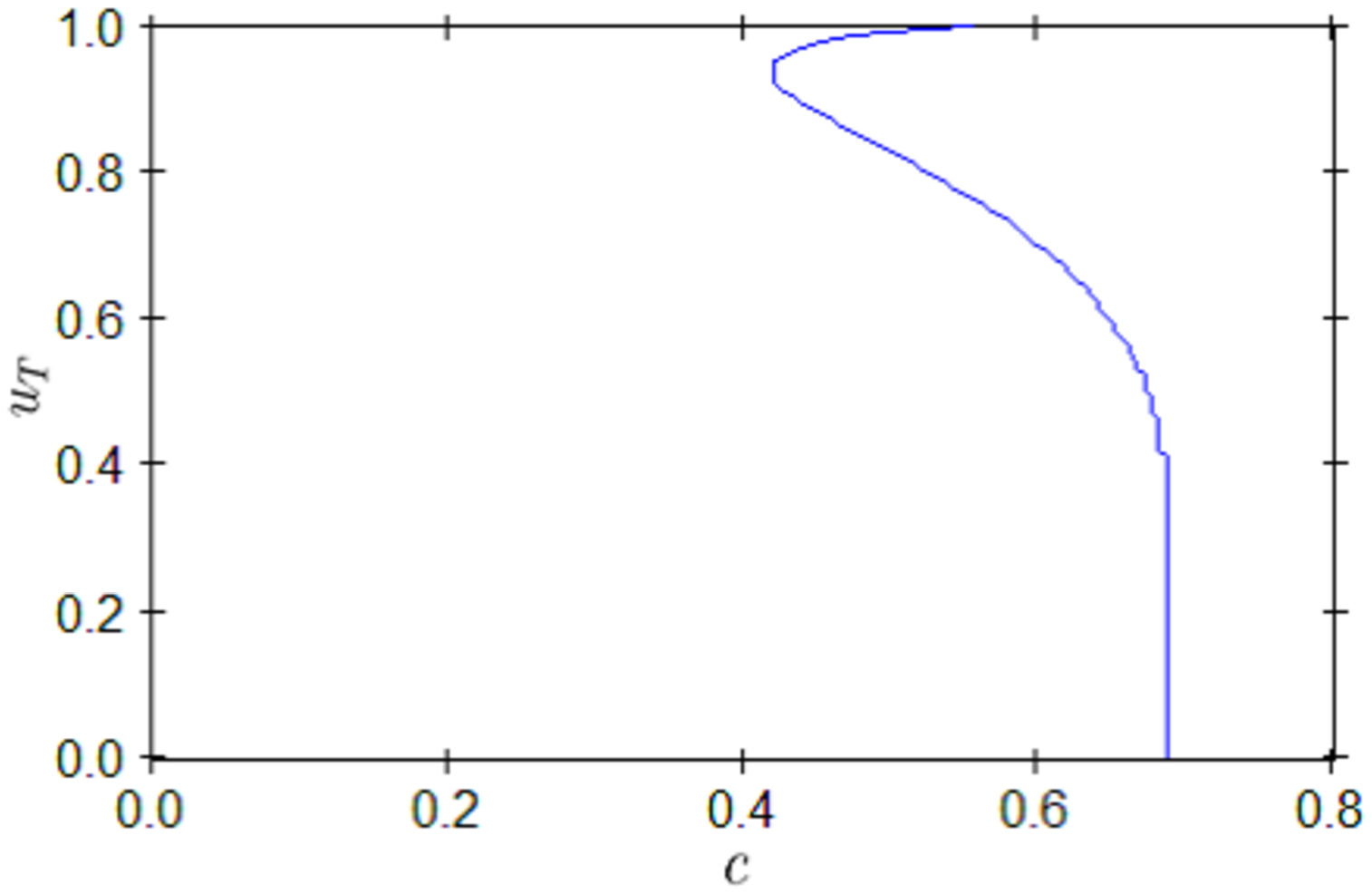}} \\
$q=0.0$ & $q=1.0$ 
\end{tabular}
\end{center}
{\bf Fig. 4}\; The phase diagram in the $(c,\;u_{T})$ plane. The area on the 
left and right side of the curve represents the phase with broken and restored 
chiral symmetry, respectively. \\[10mm]

In order to translate the phase diagrams shown in Fig. 4 in terms of the 
temperature $T$ and the chemical potential $\mu$, we rewrite $T$ and $\mu$ 
as a function with respect to the parameters $u_{T}$ and $c$. From Eq. 
(\ref{315}), the asymptotic separation $L$ between the $\text{D8-}$ and 
$\overline{\text{D8}}$-branes is obtained as a functions of $u_{T}$ and $q$\,:
\begin{align}
L=2\int^{\infty}_{U_{0}} dU \; \tau'(U) 
= \left(\dfrac{R_{\rm D4}^3}{U_{0}}\right)^{1/2} F(u_{T},\;q) \,, \label{324}
\end{align}
where 
\begin{align}
F(u_{T},\;q)
=\int^{\infty}_{1} du \dfrac{2}{\sqrt{u^3 \wt{f}(u)}}
\sqrt{\dfrac{\{1+(b-b_{0})^2\}\wt{f}(1)}{u^8\wt{f}(u)-\wt{f}(1) + (b-b_{0})^{2}
u^{3}\wt{f}(u)}} \,. \label{325}
\end{align}
The temperature $T$ given in Eq. (\ref{205}) is expressed in terms of $L$ by 
eliminating $U_{0}$\,:
\begin{align}
T=\dfrac{3}{4\pi}\left(\dfrac{u_{T}U_{0}}{R_{\rm D4}^{3}}\right)^{1/2}
= \dfrac{3}{4\pi}\dfrac{\sqrt{u_{T}}}{L}\,F(u_{T},\;q) \,. \label{326}
\end{align}
Notice that the temperature $T$ depends on not only the parameter $u_{T}$ but 
also the noncommutativity parameter $q$. 
Remember the confinement/deconfinement phase transition happens when $T=1/2\pi R$. If the ratio $L/R$ is smaller than $\dfrac{3}{2}\sqrt{u_{T}}\,F(u_{T},\;q)$, there exist the intermediate-temperature phase.
The chemical potential given in Eq. 
(\ref{321}) is also expressed in terms of $L$ as 
\begin{align}
\mu= \dfrac{R_{\rm D4}^{3}}{2\pi l_{s}^{2}L^{2}} F(u_{T},\;q)^{2} 
\int^{u}_{u_{T}}\;du'\sqrt{\dfrac{(\wt{c}+b)^{2}}{u'^{5}+(\wt{c}+b)^{2}}}\,. 
\label{327}
\end{align}
Eq. (\ref{327}) shows that the chemical potential depends on not only the 
parameters $c$ and $u_{T}$ but also the noncommutativity parameter $q$. The 
dimensionless temperature $\wt{T}$ and the dimensionless chemical potential 
$\wt{\mu}$ can be defined by using the asymptotic separation $L$ as 
\begin{align}
\wt{T} &= LT = \dfrac{3}{4\pi}\sqrt{u_{T}}\,F(u_{T},\;q)\;, 
\label{328} \\
\wt{\mu} &= \dfrac{2\pi l_{s}^{2}L^{2}}{R_{\rm D4}^{3}}\mu 
= F(u_{T},\;q)^{2} 
\int^{u}_{u_{T}}\;du'\sqrt{\dfrac{(\wt{c}+b)^{2}}{u'^{5}+(\wt{c}+b)^{2}}}
\label{329} \,. 
\end{align}
The dimensionless temperature $\wt{T}$ and the dimensionless chemical potential 
$\wt{\mu}$, as function with respect to the noncommutativity parameter $q$, are 
shown in Fig. 5 and Fig. 6, respectively. 

\clearpage

\begin{center}
\begin{tabular}{c}
\scalebox{0.8}{\includegraphics[width=100mm]{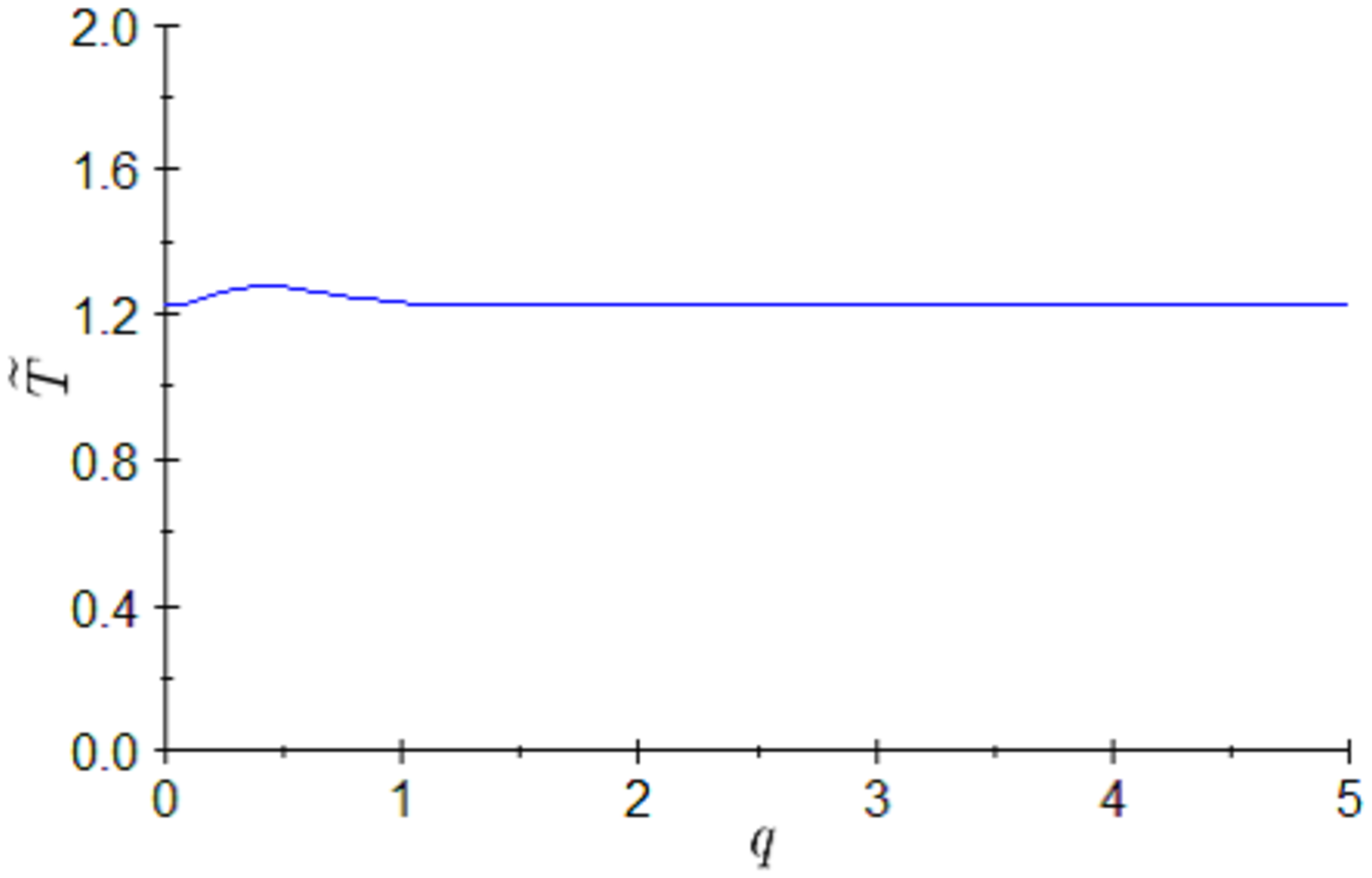}} 
\end{tabular}
\end{center}
{\bf Fig. 5}\; The dimensionless temperature $\wt{T}$ as a function of $q$ for 
$u_{T}=0.5$. The temperature is hardly affected by the noncommutativity 
parameter. \\[10mm]

\begin{center}
\begin{tabular}{cc}
\scalebox{0.8}{\includegraphics[width=100mm]{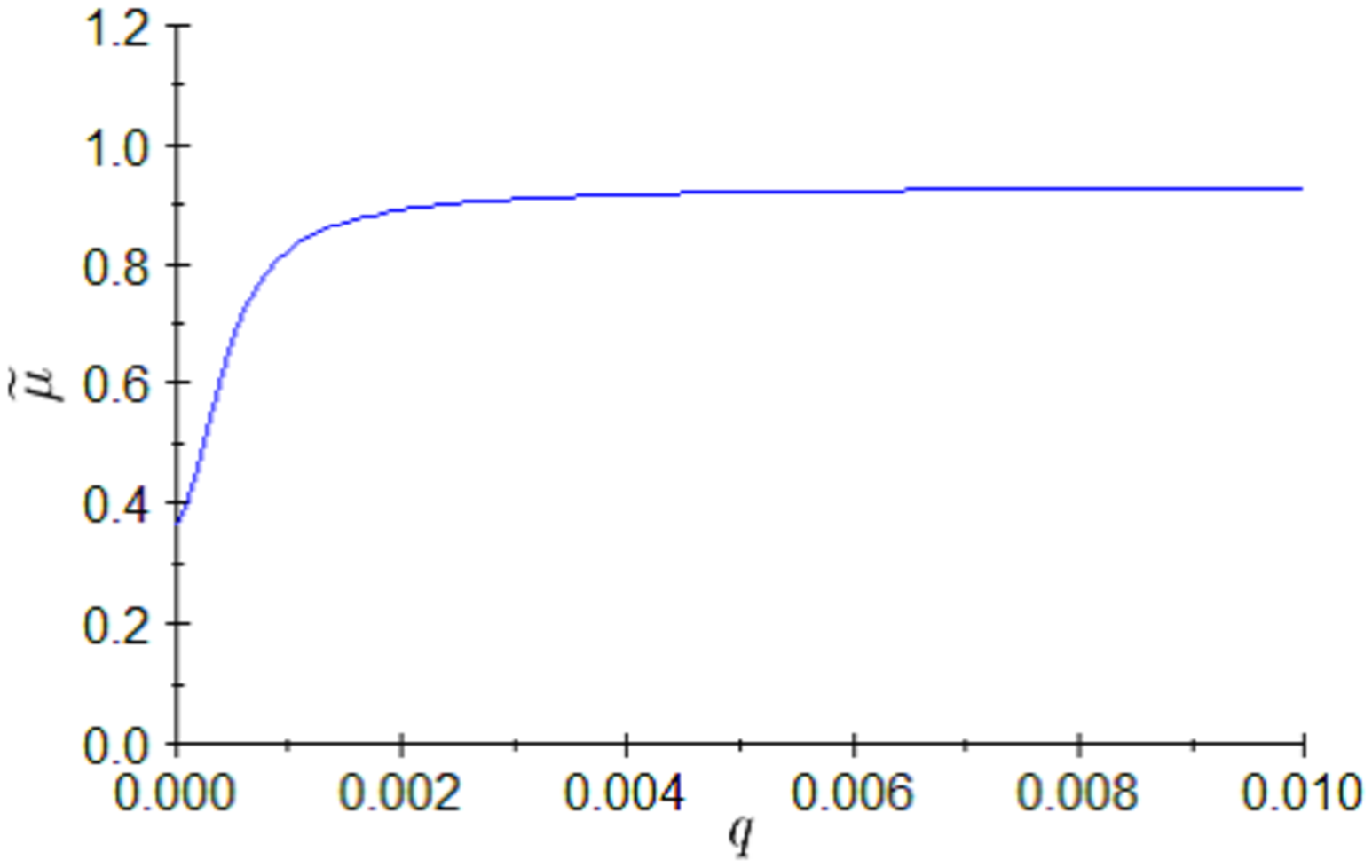}} & 
\scalebox{0.8}{\includegraphics[width=100mm]{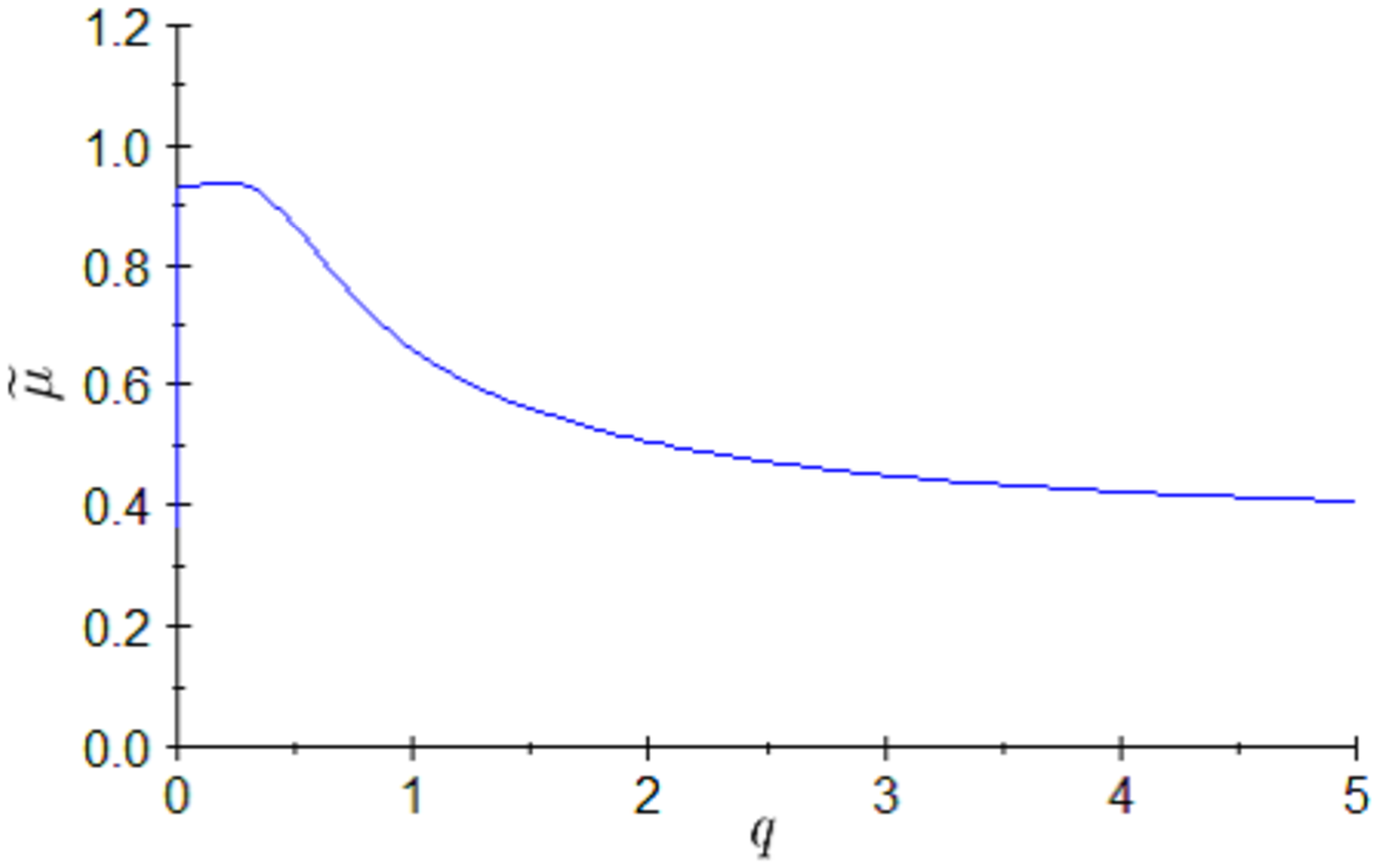}} 
\end{tabular}
\end{center}
{\bf Fig. 6}\; The dimensionless chemical potential $\wt{\mu}$ as a function of $q$ for $u_{T}=0.5$ and $c=0.5$. As $q$ increases, $\wt{\mu}$ approaches the
one in the commutative theory. \\[10mm]

Although the asymptotic separation $L$ can be regarded as approximately 
constant with fixed noncommutativity parameter, $L$ rapidly shrinks as $u_{T} 
\to 1$. The behavior of the dimensionless asymptotic separation $\wt{L}=
\left(\dfrac{U_{0}}{R_{\rm D4}^3}\right)^{1/2}L$ as a function of $u_{T}$ is 
shown in Fig. 7. 

\begin{center}
\begin{tabular}{c}
\scalebox{0.8}{\includegraphics[width=100mm]{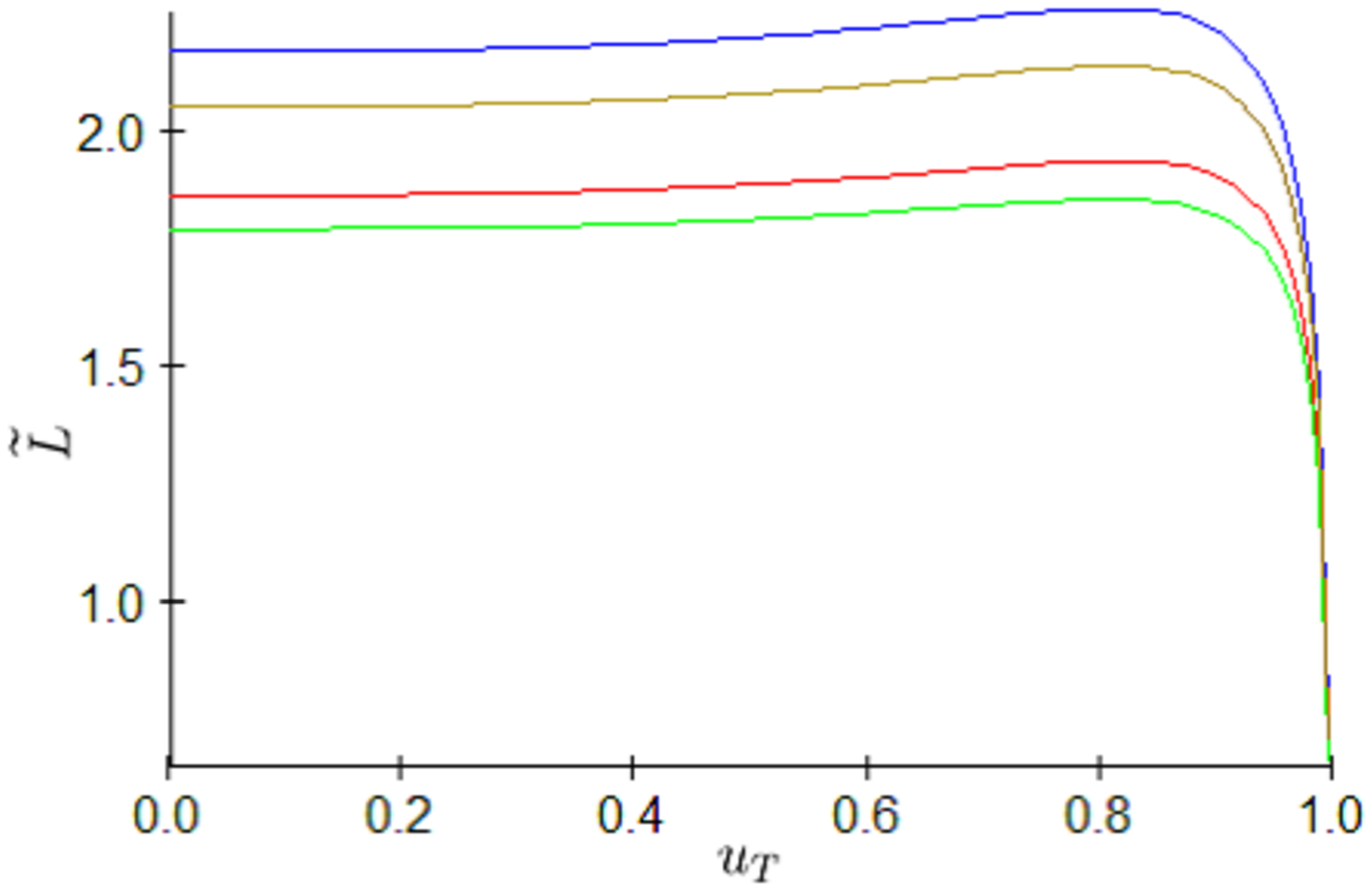}} 
\end{tabular}
\end{center}
{\bf Fig. 7}\; The dimensionless asymptotic separation $\wt{L}=
\left(\dfrac{U_{0}}{R_{\rm D4}^3}\right)^{1/2}L$ as a function of $u_{T}$ for 
some value of $q$. From the top to the bottom each line represents the value of $q$ for $0.0, \;5.0,\; 0.5, \;1.0,$ respectively. $\wt{L}$ for $q \to \infty$ 
reduces to the one in the commutative theory. \\[10mm]

By utilizing Eqs. (\ref{326}) and (\ref{327}), we can draw the phase diagram 
in terms of the temperature $T$ and the chemical potential $\mu$. The phase 
diagram in the ($\wt{\mu},\;\wt{T}$) plane is shown in Fig. 8.

\begin{center}
\begin{tabular}{cc}
\scalebox{0.8}{\includegraphics[width=100mm]{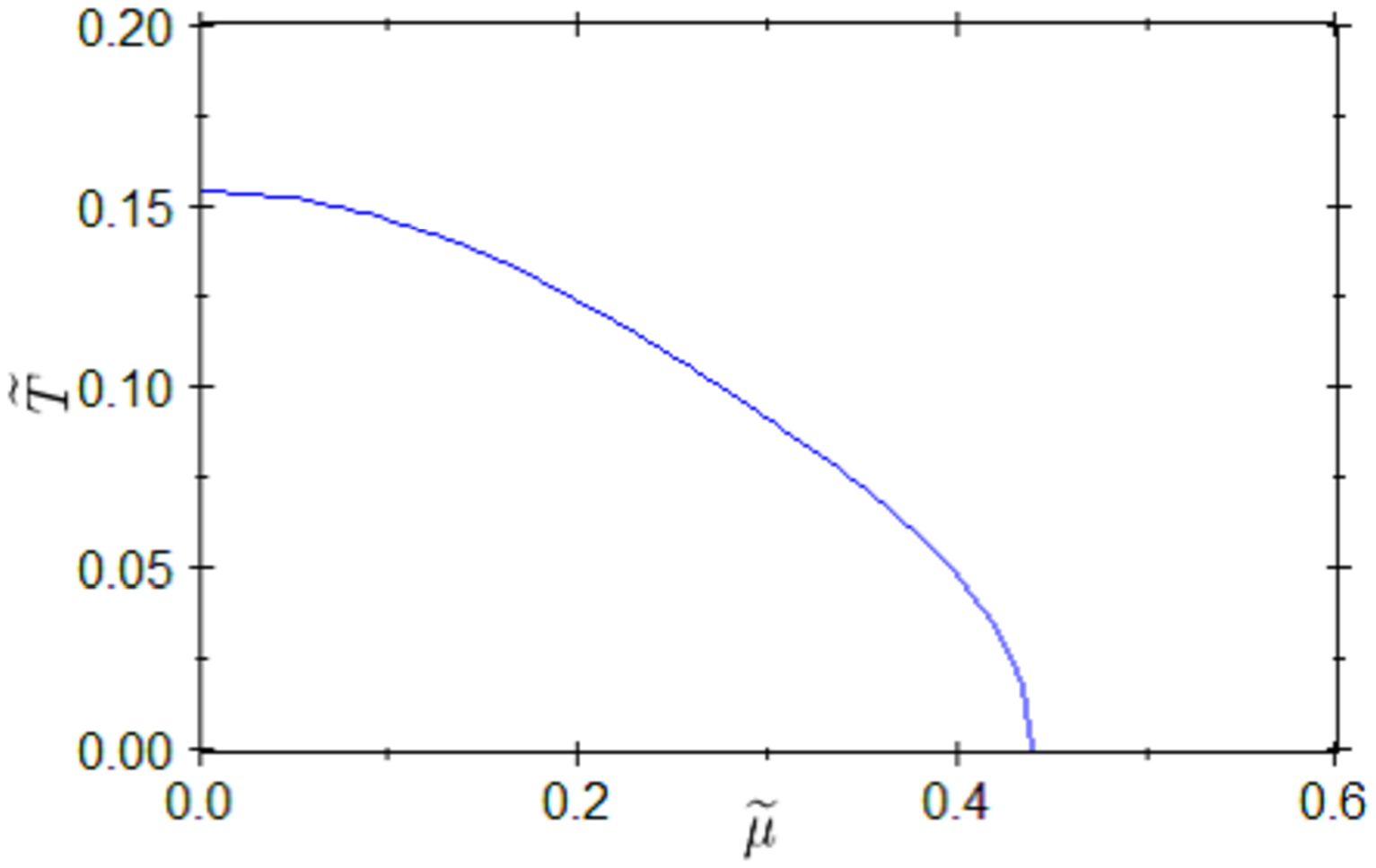}} & 
\scalebox{0.8}{\includegraphics[width=100mm]{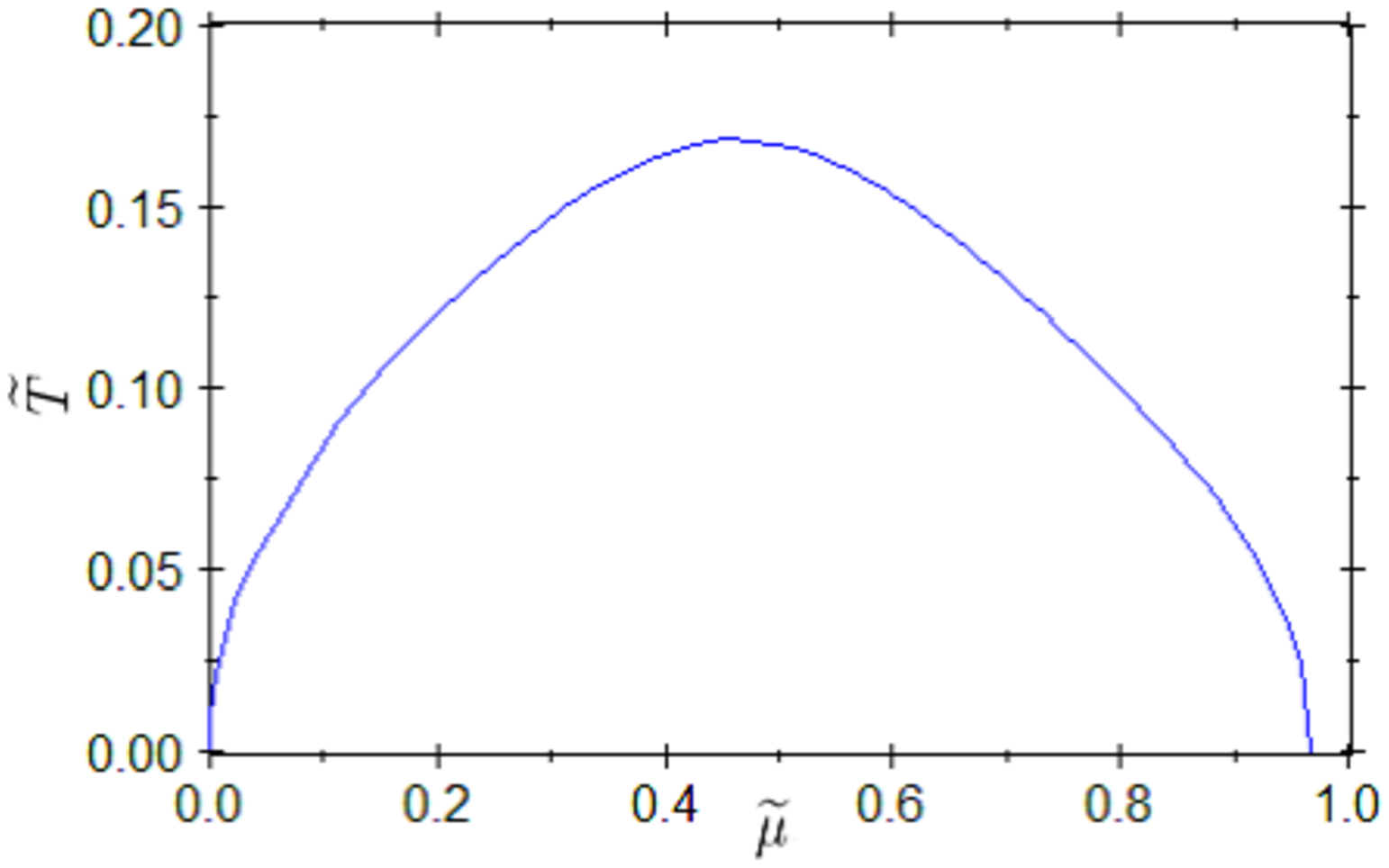}} \\
$q=0.0$ & $q=1.0$ 
\end{tabular}
\end{center}
{\bf Fig. 8}\; The phase diagram in the $(\wt{\mu},\;\wt{T})$ plane. The area 
containing the origin represents the phase with broken chiral symmetry, and 
the other area represents the phase with restored chiral symmetry. \\[10mm]

Fig. 8 indicates that the phase diagram is modified by the space 
noncommutativity. Although the chiral phase transition occurs at 
high-temperature and high-chemical potential in the noncommutative theory, the 
space noncommutativity plays a role of increasing the critical value of the 
chemical potential. The modification of the phase diagram around the origin in 
the noncommutative theory is brought about the shrinking of the asymptotic 
separation $L$.



%
%
\section{Conclusions and Discussions}
\setcounter{section}{4}
\setcounter{equation}{0}
\addtocounter{enumi}{1}

In this paper, we have constructed a noncommutative deformation of the 
holographic QCD (Sakai-Sugimoto) model after the prescription of 
Arean--Paredes--Ramallo \cite{APR}. In the same way as D3-D7 brane system, WZ 
terms in the effective action of probe $\text{D8-}$brane play the important 
role of bringing the space noncommutativity in the gauge theory side. 
%
%
Both the temperature $T$ and the chemical potential $\mu$ in the gauge theory 
side depend on the noncommutativity parameter due to the noncommutative 
deformation. Fig. 5 and 6 show that although the dependence on the 
noncommutativity parameter of the temperature is slight, that of the chemical 
potential is not very small. The critical value of the chemical potential in 
the noncommutative space tends to be larger than that in the commutative space.

The space noncommutativity has an effect on an aspect of the chiral 
phase transition for the holographic QCD at finite temperature. In general, 
intermediate temperature phase, in which the gluons deconfine but the chiral 
symmetry remains broken, is easy to be realized in some region of the 
noncommutativity parameter. The phase diagram of the noncommutative theory 
shows the phase transition occurs in ``high"-chemical potential region in 
comparison with the commutative theory. The higher critical values of the 
chemical potential are caused by the effect that space noncommutativity plays a role of increasing the critical value of the chemical potential. The fact 
indicates that the chiral symmetry restoration tends to be hard to happen in 
the noncommutative theory.

On the other hand, the phase diagram of the noncommutative theory shows the 
phase transition also occurs in low-temperature and low chemical potential 
region. It can be interpreted that the phase transition in low-temperature and 
low chemical potential region is caused by the rapidly shrinking of the 
asymptotic separation $L$, rather than the space noncommutativity. More exact 
evaluation will be required for the phase diagram of the noncommutative theory 
in low-temperature and low chemical potential region.

The magnitude of noncommutativity parameter $\theta$ denotes the degree of 
space noncommutativity in the noncommutative theory. The noncommutative QCD 
theory reduces to the commutative one when the noncommutativity parameter 
tends to zero. However, the noncommutative QCD theory also reduces to the 
commutative one in the large noncommutativity parameter limit. This property 
reminds us Morita equivalence between noncommutative tori \cite{DN}.

The noncommutative QCD tends to realize easily the intermediate temperature 
phase than ordinary QCD. In the intermediate temperature phase, there still be bound state of the hadron degrees of freedom. The property of the hadron in QCD at finite temperature can be analyzed by the holographic method. In the 
intermediate temperature phase, the masses of the low spin meson are 
temperature dependent and there is a dissociation phenomenon of the large spin 
mesons \cite{PSZ}. It is interesting to analyze the property of the hadron in 
the noncommutative QCD at finite temperature and to make clear what kind of 
influence has the noncommutativity parameter had on the property of the hadron. 

The UV/IR mixing is well known as distinctive features of noncommutative field 
theories. The UV/IR mixing appears to be the qualitative difference between 
ordinary and noncommutative field theory. The difference in chiral phase 
structure between ordinary and noncommutative QCD might be related to the UV/IR mixing. We hope to discuss this subject in the future.


\section*{Acknowledgments}

We are grateful to A. Sugamoto for useful discussions and comments. 
One of us (T.N.) would like to thank members of the Physics Depertment 
at College of Engineering, Nihon University for their encouragements. 

\clearpage

%
%

\end{document}